\documentclass[aps]{revtex4}
\usepackage[colorlinks=true, pdfstartview=FitV, linkcolor=blue, citecolor=red, urlcolor=magenta, breaklinks=true]{hyperref}
\usepackage{graphicx}  
\usepackage{amsfonts}

    \usepackage{amssymb}
    \usepackage{amsmath}
    \usepackage{subfigure}
    \usepackage{subeqnarray}

\begin{document}


\title{Exploring the thermodynamics of non-commutative scalar fields}
\author{Francisco A. Brito 
 and Elisama E.M. Lima
}
\email{fabrito@df.ufcg.edu.br, elisamafisica@gmail.com}
\affiliation{
Departamento de F\'{\i}sica, Universidade Federal de Campina Grande, Caixa Postal 10071, 58109-970 Campina Grande, Para\'{\i}ba, Brazil\\
Departamento de F\'\i sica,
Universidade Federal da Para\'\i ba, Caixa Postal 5008,
58051-970 Jo\~ ao Pessoa, Para\'\i ba,
Brazil
}
\begin{abstract} 
{ We study the thermodynamic properties of the Bose-Einstein condensate (BEC) in the context of the quantum field theory with non-commutative target space. Our main goal is to investigate in which temperature and/or energy regimes the non-commutativity can characterize some influence in the BEC properties described by a relativistic massive non-commutative boson gas. The non-commutativity parameters play a key role in the modified dispersion relations of the non-commutative fields, leading to a new phenomenology. We have obtained the condensate fraction, internal energy, pressure and specific heat of the system and taken ultra-relativistic (UR) and non-relativistic limits (NR). The non-commutative effects in the thermodynamic properties of the system are discussed. We found that there appear interesting signatures  around the critical temperature. 
}

\end{abstract}
\maketitle
\newpage
\section{Introduction}

Quantum field theories (QFT) formulated in non-commutative spaces have interesting phenomenological implications  which have been the subject of numerous investigations in recent years, in connection with many physics phenomena related to the possibility of violation of Lorentz and CPT symmetry \cite{Carroll,Hinchliffe2,Morita,Carmona2003}, analyzes of matter-antimatter asymmetry \cite{jose}, implications in cosmological models \cite{chu,Alexander,brito,li2013}, in addition to observations involving ultra-energetic cosmic rays, neutrino physics, and other sources of high energy existing in the Universe \cite{Horvat,Hinchliffe}.

Such theories, also known as non-commutative quantum field theories (NCQFT), were inspired by quantum mechanics on a non-commutative spacetime \cite {Szabo,Douglas} whose proposal is based on the idea of quantization of spacetime through the replacement of its coordinates by operators which meet the commutation relation $\left [\hat{x}^{\mu},\hat{x}^{\nu}\right]\equiv i\theta^{\mu\nu}$. This causes the spatial coordinates to become discrete and there is no longer the notion of points in the space. This proposal follows the same lines of reasoning in the usual quantum theory characterized by quantization of the phase space which results in the existence of a minimal area scaled by Planck constant.

The NCQFT was initially motivated by the interest of eliminating divergences that appear in quantum field theory, since by introducing non-commutativity is possible to exist a minimal distance in the spacetime to eliminate the infinities  \cite{SNYDER1,SNYDER2}. However, this proposal was abandoned because of the success of renormalization process.

In addition, the non-commutative Yang-Mills theory emerged later as a limit of low energy of string theory in the presence of a magnetic field background \cite{Seiberg}, causing a major boost in the resumption of the study of models involving non-commutative QFT. The reappearance of these theories is also motivated by some formulations of quantum gravity, which suggests that in regions around the Planck scale  $(l_p\sim10^{- 33}cm)$ the spacetime loses its continuous structure and the effects of quantum gravity become important \cite{DOPLICHER,Aschieri,Aschieri2}.

An important aspect presented by non-commutative quantum field theory is the violation of the Lorentz invariance, which has been the subject of great attention in QFT \cite{Carroll,Carmona2003,chaichian2011cpt}. This violation should occur only at very small distances, thus, within the limits that the non-commutative parameter tends to zero the results previously known are recovered. Theories with loss of Lorentz invariance present deformations in the dispersion relation. These deformations allow us to build phenomenological models that can be further tested in cosmology or particle physics. In this perspective, it is expected that signs of the non-commutativity appear in experiments involving cosmic microwave background, ultra-energetic cosmic rays, or other source of high energy.

The introduction of the non-commutativity in QFT can be made at the fields by modifying their canonical commutation relations \cite{Carmona2003,Camona2,jose}, which we will discuss later in the section \ref{cap:1}. These fields, called non-commutative fields, arise as a generalization of quantum mechanics in Groenewold-Moyal plane \cite{Nair}, with the target space (or space of fields) being considered as a non-commutative plane and a base space considered a commutative spacetime.

Several authors investigated the physical implications caused by these non-commutative fields. In Ref.~\cite{Carmona2003} was shown how the non-commutativity in the field space produces violation of Lorentz invariance which is manifest in the dispersion relation. Subsequently, in Ref.~\cite{balachadran}  was shown how the blackbody radiation spectrum is modified by the deformation of the canonical commutation relations of massless scalar fields. An important  aspect of this modified spectrum is that the energy density in the region of high frequency is higher than that of the usual radiation blackbody spectrum. Later, in Ref.~\cite{brito} was presented implications of this theory in inflationary cosmological  models in a universe filled with a non-commutative gas at high temperature regime. Also, in this perspective, recently in Ref.~\cite{Khelili} has been studied how the Casimir effect might be affected by the non-commutativity in the target space, where was shown the arising of a repulsive non-commutative Casimir force at the microscopic level.
 
In this paper, we consider this approach based on non-commutative fields with commutation relations in equal times given by
\begin{equation}
\left[\hat{\varphi}^{a}(\vec{x},t),\hat{\varphi}^{b}(\vec{y},t)\right]=i\epsilon^{ab}\theta(\sigma;\vec{x}-\vec{y})
\end {equation}
to investigate the Bose-Einstein condensation of a system described by non-commutative scalar fields. Here $\theta(\sigma,\vec {x} - \vec{y})$ is considered as a Gaussian distribution with a parameter $\sigma$  related to the standard deviation of the distribution. 

The Bose-Einstein condensation has some very peculiar characteristics, such as below certain temperature a finite fraction of the number of particles are condensed in the ground state. 
Theoretical and experimental studies discussing the properties of this thermodynamic phenomenon has gained great attention in different areas of physics \cite{Pethick,Anderson, Davis,Dalfovo,Harber}.

Recently several studies have been developed in the analysis of relativistic effects in Bose-Einstein condensates \cite{Grether,Sales,du,pandita} --- see also \cite{Das:2009qb} for related issues. However, it is also interesting to explore possible non-commutative consequences that become relevant for these systems at high energies. Because of this, we study the thermodynamic properties of bosonic fields described in the non-commutative quantum field theory assuming the inclusion of particle-antiparticle pair production.

The paper is organized as follows. In Sec.~\ref{cap:1} we make a revision on field theories in non-commutative spaces involving massive scalar fields, where the physical peculiarities of such theories based on the non-commutative algebra are analyzed. In Sec.~\ref{chp:2} we explore the effects due to the non-commutativity in the thermodynamic functions  of a  relativistic bosonic gas with and without anti-bosons. Expressions for these quantities were obtained, such as, particles density, internal energy, pressure and specific heat. In Sec.~\ref{chp:3} we present our results obtained through non-commutative theory which are compared to the results  established by the usual theory. Finally, in 
Sec.~\ref{chp:4} we present our final comments and conclusions.

\section{Non-commutative quantum field theory} \label{cap:1}

Field theories based on concept  of non-commutative target space is gaining more
attention and interest as matter of study, whose properties and implications caused are studied and applied to different physics problems \cite{brito,Carmona2003,Camona2,balachadran,balachadran2}. The formalism that describes non-commutative fields was proposed from an analogy with the fundamental ideas of the non-commutative quantum mechanics (NCQM), which suggests the replacement of non-commutative relationships between the coordinates of a spacetime to non-commutative relationships between the fields.

The aim of this section is to present a revision of the non-commutative quantum field theory. The concepts here formulated will be fundamental to study the thermodynamic properties of a bosonic gas with its dispersion relation modified by the introduction of new parameters related to the non-commutativity. This section is divided into two subsections. The first subsection focus on the definition of  non-commutative spacetime and the following subsection discusses an extension of this theory to a massive scalar field where the target space is considered a non-commutative space living in a commutative spacetime in (3 + 1)-dimensions.

\subsection{Non-commutative spacetime}
\label{ETNC} 

The notion of a non-commutative spacetime had already been proposed since the beginning of the quantum mechanics \cite{SNYDER1}. Motivated by the extension of the usual commutation relations between position and momentum, the NCQM imposes non-commutation relations between the coordinates of the spacetime. The coordinates $x^{\mu}$ are substituted by Hermitian operators $\hat{x}^{\mu}$ of a non-commutative algebra \footnote{Operators with the hat character indicates that they obey the non-commutative algebra.}, satisfying deformed commutation relations
\begin{subeqnarray}\label{Com}
\slabel{com1}
\left[ \hat{x}^{\mu},\hat{x}^{\nu}\right]&=&i\epsilon^{\mu\nu} \bar{\theta} \equiv 
 i\bar{\theta}^{\mu\nu}, \\
\slabel{com2}
\left[ \hat{p}_{\mu},\hat{p}_{\nu}\right]&=&0, \\
\slabel{com3}
\left[ \hat{x}^{\mu},\hat{p}_{\nu}\right]&=&i \delta^{\mu}_{\nu},
\end{subeqnarray}
where  $\bar{\theta}$ is  known as non-commutative parameter having dimension of area, and $\epsilon^{\mu\nu}$ is a real anti-symmetric matrix.

The non-commutativity of the spacetime causes effects in their geometrical structure, since the spacetime ceases to be continuous and becomes discrete, leaving aside the notion of point that is replaced by a minimal  area, called Planck cell. This implies a set of uncertainty relations among its coordinates
\begin{eqnarray}
\Delta\hat{x}^{\mu}\Delta\hat{x}^{\nu}\geq \frac{1}{2}|\bar{\theta}^{\mu\nu}|.
\end{eqnarray}
Note that in the limit $\bar{\theta}\rightarrow 0$ the usual commutative theory is obtained.

In \cite{balachadran}, see also \cite{grosse,zamo,fad}, the non-commutative coordinates are related to the usual coordinates by the deformation
\begin{subeqnarray}\label{eq:sub}
\slabel{sub1} \hat{x}^{\mu}&=&x^{\mu}-\frac{1}{2}\bar{\theta}^{\mu\nu} p_{\nu}, \\
\slabel{sub2} \hat{p}^{\mu}&=&p^{\mu},
\end{subeqnarray}

The deformation introduced above is applied to the coordinates, which are the degrees of freedom of the system. Similarly, in quantum field theory the degrees of freedom are the fields defined at every point in the spacetime, so we will introduce the non-commutativity into the fields which will engender the non-commutative target space.

\subsection{Free massive bosons description}

In this section, mainly following \cite{balachadran}, we review the theory of commutative and non-commutative massive scalar fields and then we analyze the effects caused by the non-commutativity of the fields.

\subsubsection{Commutative scalar field }

We define here the commutative target space which will be important to compare with the effects that the non-commutative fields develop on the dispersion relation. For this, consider a theory with two scalar fields in a (3 + 1)-dimensional base space and target space as a commutative plane  $\mathbb{R}^2$, then
\begin{eqnarray}
 \varphi : M_3\times \mathbb{R} &\longrightarrow &  \mathbb{R}^2 \nonumber \\
 \left( \vec{x},t\right)&\longmapsto& \varphi\left(\vec{x},t\right), \nonumber
\end{eqnarray}
where the field components are denoted by $ \varphi^{i}$ with $i=1, 2$. Let us assume that each spatial direction compactified in $S^1$ with radius $R$. This causes the field components to be periodic in the spatial coordinates $(x, y, z) $,
\begin{eqnarray}
 \varphi^{i}\left(x+R,y+R,z+R,t\right) \equiv  \varphi^{i}\left( x,y,z,t\right) \equiv  \varphi^{i}\left( \vec{x},t\right).
\end{eqnarray}
The compactification of space allows us write the components of the field $\varphi^{i}\left( \vec{x},t\right)$ as a Fourier series
\begin{eqnarray}
\label{MFourier}
 \varphi^{i}\left( \vec{x},t\right)= \sum_{\vec{n}}e^{\frac{2\pi i}{R}\vec{n}.\vec{x}} \varphi^{i}_{\vec{n}}\left(t\right),
\end{eqnarray}
where $\vec{n}=(n_1,n_2,n_3)$, \, with $n_i \in \mathbb{Z} $, and then the Fourier components are
\begin{eqnarray}
\label{MFourier1}
 \varphi^{i}_{\vec{n}}\left(t\right)=\frac{1}{R^3}\int d^3 x e^{-\frac{2\pi i}{R}\vec{n}.\vec{x}} \varphi^{i}\left( \vec{x},t\right).
\end{eqnarray}
The Lagrangian density describing massive scalars fields is given by
\begin{eqnarray}
 {\mathcal L} &=&\frac{g}{2}\sum_{i}\left[\partial_{\mu}\varphi^{i}\partial^{\mu}\varphi^{i} -  m^2({\varphi^{i}})^2\right], \nonumber \\
&=& \frac{g}{2}\sum_{i}\left[(\partial_t \varphi^{i})^2-(\nabla\varphi^{i})^2-m^2({\varphi^{i}})^2\right],
\end{eqnarray}
whose Lagrangian is $L=\int{{\mathcal L}d^3 x}$
\begin{eqnarray}
\label{Lagran1}
L&=&\frac{g}{2}\sum_{i}\int{d^3 x\left[(\partial_t \varphi^{i})^2-(\nabla\varphi^{i})^2-m^2({\varphi^{i}})^2\right]}.
\end{eqnarray}
To obtain the above Lagrangian written in terms of Fourier modes we use the components of the fields $ \varphi^{i}\left( \vec{x},t\right)$ given in (\ref{MFourier}), then
\begin{eqnarray}
\int d^3 x(\partial_t \varphi^{i})^2 &=& \sum_{\vec{n}} R^3\dot{\varphi}^{i}_{\vec{n}}\dot{\varphi}^{i}_{-\vec{n}}, \nonumber\\
\label{intg}
\int{d^3 x(\nabla\varphi^{i})^2}&=& \sum_{\vec{n}} R^3\left(\frac{2\pi|\vec{n}|}{R}\right)^2\varphi^{i}_{\vec{n}} \varphi^{i}_{-\vec{n}}, \nonumber \\
\int d^3 x(\varphi^{i})^2 &=& \sum_{\vec{n}} R^3\varphi^{i}_{\vec{n}}\varphi^{i}_{-\vec{n}}, \nonumber\\
\end{eqnarray}
resulting in
\begin{eqnarray}
\label{lagrann}
 L=\frac{gR^3}{2} \sum_{i,\vec{n}} \left\{ \dot{\varphi}^{i}_{\vec{n}} \dot{\varphi}^{i}_{-\vec{n}}-  \left[\left(\frac{2\pi|\vec{n}|}{R}\right)^2 +m^2\right]\varphi^{i}_{\vec{n}} \varphi^{i}_{-\vec{n}}\right\}.
\end{eqnarray}
The canonical momenta associated with the Fourier modes $\dot{\varphi}^{i}_{\vec{n}}$ \, are
\begin{eqnarray}
 \pi^{i}_{\vec{n}} &=& \frac{\partial L}{\partial \dot{\varphi}^{i}_{\vec{n}}} \nonumber \\
&=&  \frac{gR^3}{2}\sum_{j,\vec{m}}\frac{\partial}{\partial \dot{\varphi}^{i}_{\vec{n}}} (\dot{\varphi}^{j}_{\vec{m}}\dot{\varphi}^{j}_{-\vec{m}}) \nonumber \\
&=& \frac{gR^3}{2}\sum_{j,\vec{m}}\left[\delta^{j}_{i}\delta^{\vec{m}}_{\vec{n}}\dot{\varphi}^{j}_{-\vec{m}}+ \dot{\varphi}^{j}_{\vec{m}} \delta^{j}_{i}\delta^{-\vec{m}}_{\vec{n}})\right]\nonumber \\
&=&gR^3\dot{\varphi}^{i}_{-\vec{n}}.
\end{eqnarray}
Now using the Lagrangian  (\ref{lagrann}) and applying a Legendre transform, we obtain the Hamiltonian $H$ as a function of the modes $\varphi^{i}_{\vec{n}}$ and canonical momenta $ \pi^{i}_{\vec{n}}$
\begin{eqnarray}
\label{Hami1}
 H &=&  \sum_{i,\vec{n}}\pi_{\vec{n}}^{i}\dot{\varphi}^{i}_{\vec{n}} - L, \nonumber \\
&=&  \sum_{i,\vec{n}} \left\{\frac{ \pi_{\vec{n}}^{i}\pi_{-\vec{n}}^{i}}{2gR^3}
   + \frac{gR^3}{2}\left[\left(\frac{2\pi|\vec{n}|}{R}\right)^2 +m^2\right]\varphi^{i}_{\vec{n}} \varphi^{i}_{-\vec{n}}\right\}, \nonumber\\
\label{Hami3}
&=&	 \sum_{i,\vec{n}} \left\{\frac{ \pi_{\vec{n}}^{i}\pi_{-\vec{n}}^{i}}{2gR^3}
   + \frac{gR^3}{2}\omega_{\vec{n}}^2\varphi^{i}_{\vec{n}} \varphi^{i}_{-\vec{n}}\right\}.
\end{eqnarray}
This Hamiltonian is equivalent to an infinite set of uncoupled harmonic oscillators with frequencies
\begin{equation}
\omega_{\vec{n}}=\sqrt{\left(\frac{2\pi|\vec{n}|}{R}\right)^2 + m^2}.
\end{equation}

\subsubsection{Non-commutative scalar field}
\label{EANC}

In this subsection, the commutative plane defined above is substituted by a non-commutative plane in $\mathbb{R}^2$. The way to introduce non-commutativity in the fields space is analogous to the deformation  made on the spacetime coordinates presented in section (\ref{ETNC}). The scalar field with non-commutative target space $\mathbb{R}^2$, denoted by $\hat{\varphi}^{a}(\vec{x},t)$, can be written in terms of commutative fields as \cite{balachadran} 
\begin{subeqnarray}\label{eq1NC}
\hat{\varphi}^{a}(\vec{x},t)&=&\varphi^{a}(\vec{x},t)-\frac{1}{2}\epsilon^{ab}\theta\pi_{b}(\vec{x},t),\\
\hat{\pi}_{a}(\vec{x},t)&=&\pi_{a}(\vec{x},t).
\end{subeqnarray}
Thus, the commutation relations in equal times for non-commutative scalar fields introduced above
become (see appendix \ref{appendixA})
\begin{subeqnarray} \label{cc}
 \slabel{comutador1} \left[ \hat{\varphi}^{a}(\vec{x},t),\hat{\varphi}^{b}(\vec{y},t)\right]&=&i\epsilon^{ab}\theta\delta(\vec{x}-\vec{y}), \\
 \left[ \hat{\pi}_{a}(\vec{x},t),\hat{\pi}_{b}(\vec{y},t)\right]&=&0,\\
 \left[ \hat{\varphi}^{a}(\vec{x},t),\hat{\pi}_{b}(\vec{y},t)\right]&=&i\delta_{b}^{a}\delta(\vec{x}-\vec{y}).
\end{subeqnarray}
In this scenario  $\theta$  is the non-commutative parameter with dimension of length.

\begin{figure}[!h]
	\centering
		\includegraphics[width=7.5cm,height=7.5cm]{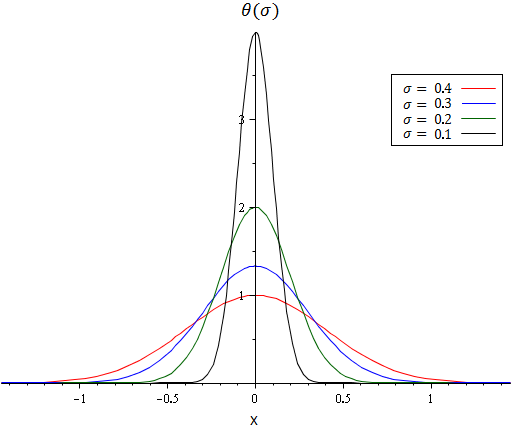}
	\caption[The function $\theta (\sigma)$ for different values of $\sigma $.] {The function $\theta (\sigma) $ for different values of $\sigma$, with $\theta = 1.0$.}
\label{delta2}
\end{figure}

The function $\theta\delta(\vec{x}-\vec{y})$ that appears in Eq.~(\ref{comutador1}) has been regularized by a Gaussian-type distribution \cite{balachadran}, written in terms of a new parameter $\sigma$ associated with the standard deviation of the distribution
\begin{eqnarray}
\label{delta}
 \theta(\sigma)=\frac{\theta}{(\sqrt{2\pi}\sigma)^3}\mbox{exp}\left[-\sum_{i=1}^{3}\frac{(x_{i}-y_{i})^2}{2\sigma^2}\right],
\end{eqnarray}
where $\theta(\sigma)\equiv \theta(\sigma;\vec{x}-\vec{y}) = \theta\delta(\vec{x}-\vec{y})$ in the limit $\sigma \rightarrow 0$, $\theta(\sigma)$ has dimension of  $(\mbox{length})^{-2}$, and we assume the simplification 
$\sigma_1=\sigma_2=\sigma_3=\sigma$ for the argument of the exponential. The behaviour of this function $\theta (\sigma)$ can still be seen in the Figure [\ref{delta2}] for different values of $\sigma$.

We note that the regularization parameter $\sigma$ controls the width of the Gaussian describing the range of the non-commutativity, that is, in the limit as this parameter goes to zero we get the usual Dirac delta function and in the limit as the parameter grows we recover the commutativity  of the fields. This regularization proved to be useful in study of the radiation spectrum of deformed black body \cite{balachadran}, where it was predicted a divergence in the energy density with respect to frequency increase when $\sigma=0$  and that the deviation of the new radiation spectrum of the deformed blackbody has a stronger dependence on  $\sigma$ than $\theta$. Besides, $\theta(\sigma)$ can be written in terms of
\begin{eqnarray}
\theta (n)=\theta e^{-\frac{2\pi^2\sigma^2|\vec{n}|^2}{R^2}}.
\end{eqnarray}
Analogous to Eq.~(\ref{MFourier}) the non-commutative fields $\hat{\varphi}^{a}(\vec{x},t)$ can be written in terms of Fourier modes
\begin{eqnarray}
\label{FourierNC}
 \hat{\varphi}^{a}\left( \vec{x},t\right)= \sum_{\vec{n}}e^{\frac{2\pi i}{R}\vec{n}.\vec{x}} \hat{\varphi}^{a}_{\vec{n}}\left(t\right).
\end{eqnarray}
Then, using equations (\ref{eq1NC}) we can get dressed transformations 
\begin{subeqnarray} \label{eq2NC}
\hat{\varphi}^{a}_{\vec{n}}&=&\varphi^{a}_{\vec{n}}-\frac{1}{2R^3}\epsilon^{ab}\theta(n)\pi^{b}_{-{\vec{n}}}, \\
\hat{\pi}^{a}_{\vec{n}}&=&\pi^{a}_{\vec{n}}.
\end{subeqnarray}
%
The commutation relations for the Fourier components become (appendix \ref{appendixA})
\begin{subeqnarray} \label{comutador2}
\left[\hat{\varphi}^{a}_{\vec{n}},\hat{\varphi}^{b}_{\vec{m}}\right] &=& \frac{i\epsilon^{ab}\theta(n)}{R^3}\delta_{\vec{n}+\vec{m},0},\\
\left[\hat{\pi}^{a}_{\vec{n}},\hat{\pi}^{b}_{\vec{m}}\right] &=&0, \\
\left[\hat{\varphi}^{a}_{\vec{n}},\hat{\pi}^{b}_{\vec{m}}\right] &=& i\delta^{ab}\delta_{\vec{n},\vec{m}}.
\end{subeqnarray}
The non-commutativity of the fields introduced in (\ref{cc}) is equivalent to replace the Hamiltonian operator (\ref{Hami3}) for a new non-commutative Hamiltonian
\begin{eqnarray}
\label{Hami1NC}
H_{NC}&=& \sum_{i,\vec{n}}\left[\frac{\hat{\pi}^{i}_{\vec{n}}\hat{\pi}^{i}_{-\vec{n}}}{2gR^3} + 
\frac{gR^3}{2}\omega_{\vec{n}}^{2}\hat{\varphi}^{i}_{\vec{n}}\hat{\varphi}^{i}_{-\vec{n}}\right].
\end{eqnarray}
Using the equations (\ref{eq2NC}) we find
\begin{eqnarray}
\label{Hami2NC}
H_{NC}&=& \sum_{i,\vec{n}}\left[ \frac{\Omega_{\vec{n}}^{2}}{2gR^3}\pi^{i}_{\vec{n}}\pi^{i}_{-\vec{n}} + \frac{gR^3}{2}\omega_{\vec{n}}^{2}\varphi^{i}_{\vec{n}}\varphi^{i}_{-\vec{n}} - \frac{g}{2}\omega_{\vec{n}}^{2}\theta(n)\epsilon_{ik}\varphi^{i}_{\vec{n}}\pi^{k}_{\vec{n}}\right],
\end{eqnarray}
where
\begin{eqnarray}
\Omega_{\vec{n}}^{2}=1+\left(\frac{ g \omega_{\vec{n}}\theta(n)}{2}\right)^2 \,\,\,\,\,\,\,\,\,\,\,\mbox{and} \,\,\,\,\,\,\,\,\,\,\, \omega_{\vec{n}}=\sqrt{\left(\frac{2\pi|\vec{n}|}{R}\right)^2 + m^2}.
\end{eqnarray}
Note from  Eq.~(\ref{Hami2NC}) that the dispersion relation for these boson fields was modified. If we take the limit $\theta \rightarrow 0$ we see that $\theta(n)\rightarrow 0$ and $\Omega_{\vec{n}}^{2}=1$, falling into the usual dispersion relation, Eq.~(\ref{Hami3}).

It is remarkable that the Hamiltonian (\ref{Hami2NC}) has an equivalent term to a set of infinite harmonic oscillators and another proportional to the $z$-component of the angular momentum, i.e.,
\begin{subeqnarray} \label{hammi}
\slabel{Hami4NC}
H_{\vec{n}}&=&\sum_{i}\left[ \frac{\Omega_{\vec{n}}^{2}}{2gR^3}\pi^{i}_{\vec{n}}\pi^{i}_{-\vec{n}} + \frac{gR^3}{2}\omega_{\vec{n}}^{2}\varphi^{i}_{\vec{n}}\varphi^{i}_{-\vec{n}}\right]\,\,\,\,\, \mbox{(harmonic oscillators)},\\
\slabel{momento}
J_{\vec{n}}^{z}&=&\sum_{i,k}\epsilon_{ik}\varphi^{i}_{\vec{n}}\pi^{k}_{\vec{n}} \,\,\,\,\,\,\,\,\,\,\,\,\,\,\,\,\,\,\,\,\,\,\,\,\,\,\,\,\,\,\,\,\,\,\,\,\,\,\,\,\,\,\,\,\,\,\,\,\,\,\,\,\,\,\,\,\,\,\,\,\,\,\,\,\,\,\,\,\mbox{(angular momentum).}
\end{subeqnarray} 
Thus, 
\begin{eqnarray}
\label{Hami3NC}
H_{NC}&=& \sum_{\vec{n}}\left[ H_{\vec{n}}- \frac{g}{2}\omega_{\vec{n}}^{2}\theta(n)J_{\vec{n}}^{z}\right].
\end{eqnarray}
In addition, $H_{\vec{n}}$ given by (\ref{Hami4NC}) can be written in the usual form of the Hamiltonian of a harmonic oscillator
\begin{eqnarray}
H_{\vec{n}}&=& \sum_{i}\left[ \frac{1}{2M}\pi^{i}_{\vec{n}}\pi^{i}_{-\vec{n}} + \frac{1}{2}M\bar{\omega}_{\vec{n}}^{2}\varphi^{i}_{\vec{n}}\varphi^{i}_{-\vec{n}}\right],
\end{eqnarray}
with frequency $\bar{\omega}_{\vec{n}}$ and mass $M$ given by
\begin{equation}
\bar{\omega}_{\vec{n}}=\Omega_{\vec{n}}\omega_{\vec{n}}\,\,\,\,\,\,\,\,\,\,\,\, \mbox{and} \,\,\,\,\,\,\,\,\,\,\,\, M=\frac{gR^3}{\Omega_{\vec{n}}^2}.
\end{equation}
The hermiticity of the fields $\varphi^{i}$ implies that
\begin{subeqnarray}\label{Adjunto}
\slabel{ad1} {\varphi^{i}_{\vec{n}}}^{\dagger}(t)&=&\frac{1}{R^3}\int d^3 x e^{\frac{2\pi i}{R}\vec{n}.\vec{x}} \varphi^{i}\left( \vec{x},t\right) 
 = \varphi^{i}_{-\vec{n}}(t), \\
\slabel{ad2}  {\pi^{i}_{\vec{n}}}^{\dagger}&=& gR^{3}({\dot{\varphi}^{i}_{-\vec{n}}})^{\dagger} = gR^{3}\dot{\varphi}^{i}_{\vec{n}}  = \pi^{i}_{-\vec{n}}.
\end{subeqnarray}
So we can diagonalize $H_{\vec{n}}$ by the introduction of creation and annihilation operators
\begin{subeqnarray}\label{crianiqui}
\slabel{CA}
a_{\vec{n}}^{i}&=&\sqrt{\frac{\Delta_{\vec{n}}}{2}}\left( \varphi^{i}_{\vec{n}} + i\frac{\pi^{i}_{-\vec{n}}}{\Delta_{\vec{n}}} \right),\\
\slabel{CAA}
{a_{\vec{n}}^{i}}^{\dagger}&=&\sqrt{\frac{\Delta_{\vec{n}}}{2}}\left( \varphi^{i}_{-\vec{n}} - i\frac{\pi^{i}_{\vec{n}}}{\Delta_{\vec{n}}} \right),
\end{subeqnarray}
where
\begin{eqnarray}
\Delta_{\vec{n}}=M\bar{\omega}_{\vec{n}}=\frac{gR^3\omega_{\vec{n}}}{\Omega_{\vec{n}}}.
\end{eqnarray}
The commutation relations for ${a_{\vec{n}}^{i}}^{\dagger}$ and $a_{\vec{n}}^{i}$ are (appendix \ref{appendixA})
\begin{subeqnarray}
\label{comutador3}
\left[a^{i}_{\vec{m}},a^{j}_{\vec{n}}\right] &=& 0,\\
\left[{a^{i}_{\vec{m}}}^\dagger,{a^{j}_{\vec{n}}}^\dagger\right] &=& 0,\\
\left[a^{i}_{\vec{m}},{a^{j}_{\vec{n}}}^\dagger\right] &=& \delta^{ij}\delta_{\vec{m},\vec{n}}.
\end{subeqnarray}
The fields in terms of creation and annihilation operators, Eq.~(\ref{crianiqui}), become
\begin{subeqnarray} \label{camps}
\slabel{campos1}
\varphi_{\vec{n}}^{i}&=&\frac{1}{\sqrt{2\Delta_{\vec{n}}}}\left(a^{i}_{\vec{n}} + {a^{i}_{-\vec{n}}}^\dagger \right), \\
\slabel{campos2}
\pi^{i}_{\vec{n}}&=&-i\sqrt{\frac{\Delta_{\vec{n}}}{2}}\left(a^{i}_{-\vec{n}} - {a^{i}_{\vec{n}}}^\dagger \right).
\end{subeqnarray}
Substituting Eqs.~(\ref{camps}) in Eq.~(\ref{Hami4NC}) we get $H_{\vec{n}}$ in terms of creation and annihilation operators
\begin{equation}
\label{Hami6NC}
H_{\vec{n}}=\sum_{i}\frac{\Omega_{\vec{n}}\omega_{\vec{n}}}{2}\left(a^{i}_{\vec{n}}{a^{i}_{\vec{n}}}^\dagger + {a^{i}_{-\vec{n}}}^\dagger a^{i}_{-\vec{n}} \right).
\end{equation}
We can still write the operators in the normal order (represented by the symbol : :) by rearranging the creation and annihilation operators through the use of the  algebra, that is
\begin{eqnarray}
\sum_{\vec{n}} H_{\vec{n}} &=& \sum_{i,\vec{n}} \frac{\Omega_{\vec{n}}\omega_{\vec{n}}}{2} :\left(a^{i}_{\vec{n}}{a^{i}_{\vec{n}}}^\dagger + {a^{i}_{-\vec{n}}}^\dagger a^{i}_{-\vec{n}} \right): \nonumber \\
&=&\sum_{i,\vec{n}}  \frac{\Omega_{\vec{n}}\omega_{\vec{n}}}{2} \left({a^{i}_{\vec{n}}}^\dagger a^{i}_{\vec{n}} + {a^{i}_{-\vec{n}}}^\dagger a^{i}_{-\vec{n}} \right) \nonumber\\
&=&\sum_{i,\vec{n}} \Omega_{\vec{n}}\omega_{\vec{n}} {a^{i}_{\vec{n}}}^\dagger a^{i}_{\vec{n}}.
\end{eqnarray}
For the term of angular momentum, Eq.(\ref{momento}), we have
\begin{eqnarray}
\sum_{\vec{n}} J_{\vec{n}}^{z}&=&\sum_{i,k,\vec{n}}\epsilon_{ik}\varphi^{i}_{\vec{n}}\pi^{k}_{\vec{n}}, \nonumber \\
&=&-i \sum_{i,k,\vec{n}} \epsilon_{ik} {a^{i}_{\vec{n}}}^\dagger a^{k}_{\vec{n}}.
\end{eqnarray}
Substituting these results into Eq.~(\ref{Hami3NC}) the Hamiltonian becomes
\begin{eqnarray}
\label{Hami7NC}
H_{NC}&=&\sum_{i,\vec{n}}\left[\Omega_{\vec{n}}\omega_{\vec{n}}{a^{i}_{\vec{n}}}^\dagger a^{i}_{\vec{n}} + i\frac{g}{2}\omega_{\vec{n}}^{2}\theta(n)\epsilon_{ik} {a^{i}_{\vec{n}}}^\dagger a^{k}_{\vec{n}}\right],
\end{eqnarray}
or
\begin{eqnarray}
\label{Hami8NC}
H_{NC}&=&\sum_{\vec{n}}\left[\Omega_{\vec{n}}\omega_{\vec{n}}\left({a^{1}_{\vec{n}}}^\dagger a^{1}_{\vec{n}} + {a^{2}_{\vec{n}}}^\dagger a^{2}_{\vec{n}}  \right) + i\frac{g}{2}\omega_{\vec{n}}^{2}\theta(n)\left({a^{1}_{\vec{n}}}^\dagger a^{2}_{\vec{n}} - {a^{2}_{\vec{n}}}^\dagger a^{1}_{\vec{n}} \right)\right].
\end{eqnarray}

Let us now define new  creation ${A_{\vec{n}}^{i}}^\dagger$ and annihilation $A_{\vec{n}}^{i}$ operators as
\begin{subeqnarray}
\label{oper1}
A_{\vec{n}}^{1}&=&\frac{1}{\sqrt{2}}\left(a^{1}_{\vec{n}} -i a^{2}_{\vec{n}} \right) \,\,\,\,\,\,\,\,\,\,\,\,\, \Rightarrow \,\,\,\,\,\,\,\,\,\,\,\,\,
{A_{\vec{n}}^{1}}^{\dagger}=\frac{1}{\sqrt{2}}\left({a^{1}_{\vec{n}}}^\dagger + i {a^{2}_{\vec{n}}}^\dagger \right), \\
\label{oper2}
A_{\vec{n}}^{2}&=&\frac{1}{\sqrt{2}}\left(a^{1}_{\vec{n}} + i a^{2}_{\vec{n}} \right) \,\,\,\,\,\,\,\,\,\,\,\,\, \Rightarrow \,\,\,\,\,\,\,\,\,\,\,\,\,
{A_{\vec{n}}^{2}}^{\dagger}=\frac{1}{\sqrt{2}}\left({a^{1}_{\vec{n}}}^\dagger - i {a^{2}_{\vec{n}}}^\dagger \right), 
\end{subeqnarray}
thus
\begin{subeqnarray}\label{aniqui}
\slabel{aniq1}
a_{\vec{n}}^{1}&=&\frac{1}{\sqrt{2}}\left(A^{1}_{\vec{n}} + A^{2}_{\vec{n}} \right) \,\,\,\,\,\,\,\,\,\,\,\,\, \,\,\,\Rightarrow \,\,\,\,\,\,\,\,\,\,\,\,\,\,\,\,
{a_{\vec{n}}^{1}}^\dagger = \frac{1}{\sqrt{2}}\left({A^{1}_{\vec{n}}}^\dagger + {A^{2}_{\vec{n}}}^\dagger \right), \\ 
\slabel{aniq2}
a_{\vec{n}}^{2}&=& i\frac{1}{\sqrt{2}}\left(A^{1}_{\vec{n}} - A^{2}_{\vec{n}} \right) \,\,\,\,\,\,\,\,\,\,\,\,\,\, \Rightarrow \,\,\,\,\,\,\,\,\,\,\,\,\,\,
{a_{\vec{n}}^{2}}^\dagger = -i\frac{1}{\sqrt{2}}\left({A^{1}_{\vec{n}}}^\dagger - {A^{2}_{\vec{n}}}^\dagger \right). 
\end{subeqnarray}
The components of the fields and momenta, Eqs.~(\ref{camps}), in terms of these new operators are
\begin{subeqnarray}
\varphi_{\vec{n}}^{1}&=&\frac{1}{2\sqrt{\Delta_{\vec{n}}}}\left(A^{1}_{\vec{n}} + A^{2}_{\vec{n}} + {A^{1}_{-\vec{n}}}^\dagger + {A^{2}_{-\vec{n}}}^\dagger \right), \\
\varphi_{\vec{n}}^{2}&=&\frac{i}{2\sqrt{\Delta_{\vec{n}}}}\left(A^{1}_{\vec{n}} - A^{2}_{\vec{n}} - {A^{1}_{-\vec{n}}}^\dagger + {A^{2}_{-\vec{n}}}^\dagger \right), \\
\pi_{\vec{n}}^{1}&=&\frac{i\sqrt{\Delta_{\vec{n}}}}{2}\left({A^{1}_{\vec{n}}}^\dagger - A^{1}_{-\vec{n}}
+{A^{2}_{\vec{n}}}^\dagger - A^{2}_{-\vec{n}} \right), \\
\pi_{\vec{n}}^{2}&=&\frac{\sqrt{\Delta_{\vec{n}}}}{2}\left( {A^{1}_{\vec{n}}}^\dagger + A^{1}_{-\vec{n}}
- A^{2}_{-\vec{n}}- {A^{2}_{\vec{n}}}^\dagger \right).
\end{subeqnarray}
The operators ${A_{\vec{n}}^{i}}^\dagger$ and $A_{\vec{n}}^{i}$ satisfy the following commutation relations
\begin{eqnarray}
\label{comutador4}
\left[A^{i}_{\vec{m}},A^{j}_{\vec{n}}\right] = \left[{A^{i}_{\vec{m}}}^\dagger,{A^{j}_{\vec{n}}}^\dagger\right]=0,
\,\,\,\,\,\,\,\,\,\,\,\,\,\,\,\,\,\,\,\,
\left[A^{i}_{\vec{m}},{A^{j}_{\vec{n}}}^\dagger\right] &=& \delta^{ij}\delta_{\vec{m},\vec{n}}.
\end{eqnarray}
Substituting Eqs.~(\ref{aniqui}) into Eq.~(\ref{Hami8NC}) we obtain the Hamiltonian in terms of ${A_{\vec{n}}^{i}}^\dagger$ and $A_{\vec{n}}^{i}$
\begin{eqnarray}
H_{NC}&=& \sum_{\vec{n} }\left\{\omega_{\vec{n}} \Omega_{\vec{n}}({A_{\vec{n}}^{1}}^\dag A_{\vec{n}}^{1} + {A_{\vec{n}}^{2}}^\dag A_{\vec{n}}^{2}) - \frac{g\omega_{\vec{n}}^2\theta(n)}{2} ({A_{\vec{n}}^{1}}^\dag A_{\vec{n}}^{1} - {A_{\vec{n}}^{2}}^\dag A_{\vec{n}}^{2}) \right\}, \nonumber\\
&=&  \sum_{\vec{n} } \omega_{\vec{n}}\left\{\left(\Omega_{\vec{n}} - \frac{g\omega_{\vec{n}}\theta(n)}{2}\right){A_{\vec{n}}^{1}}^\dag A_{\vec{n}}^{1} +\left(\Omega_{\vec{n}} + \frac{g\omega_{\vec{n}}\theta(n)}{2}\right){A_{\vec{n}}^{2}}^\dag A_{\vec{n}}^{2}  \right\}. 
\label{hamil1}
\end{eqnarray}
That can be written as
\begin{eqnarray}
\label{hamil2}
H_{NC}&=& \sum_{\vec{n}}\omega_{\vec{n}}\left[\Lambda_{\vec{n}}^{1} N_{\vec{n}}^{1}  + \Lambda_{\vec{n}}^{2} N_{\vec{n}}^{2}\right],
\end{eqnarray}
where we defined the quantum numbers $N_{\vec{n}}^{1}={A_{\vec{n}}^{1}}^\dag A_{\vec{n}}^{1}$ and $N_{\vec{n}}^{2} ={A_{\vec{n}}^{2}}^\dag A_{\vec{n}}^{2}$, besides the terms of energy deformation
\begin{subeqnarray}\label{eqs-45-energ}
\Lambda_{\vec{n}}^{1}&=&\sqrt{1+\left(\frac{ g \omega_{\vec{n}}\theta(n)}{2}\right)^2}-\frac{g\omega_{\vec{n}}\theta(n)}{2}, \\
\Lambda_{\vec{n}}^{2}&=&\sqrt{1+\left(\frac{ g \omega_{\vec{n}}\theta(n)}{2}\right)^2}+\frac{g\omega_{\vec{n}}\theta(n)}{2}.
\end{subeqnarray}
We can see that the free theory of the non-commutative scalar field with two real scalar components leads to the emergence of two types of particles, which correspond to particle and antiparticle that are no more degenerated \cite{Camona2}. There are small corrections, parametrized by $\theta$ and $\sigma$, on their standard energy expression $E=\sqrt{k^2+m^2}$, where $k=2\pi|\vec{n}|/R$ is the  wave number.

Taking the energy relations $E^{1}_{\vec{k}}=\omega_{\vec{k}}\Lambda^{1}_{\vec{k}}$ for particles and $E^{2}_{\vec{k}}=\omega_{\vec{k}}\Lambda^{2}_{\vec{k}}$ for antiparticles descriptions, we consider $N_{\vec{k}}^{1}$ and $N_{\vec{k}}^{2}$ being the number operators of bosons and antibosons with momentum $k$, respectively. This enables us to analyze the thermodynamic properties of a charged and uncharged non-commutative bosonic gas. 
\section{Thermodynamics of the non-commutative fields}\label{chp:2}

In the previous section we explored the implications  that the non-commutativity of the target space causes in the dispersion relation of a system constituted by massive scalar bosons. In this section we will formulate the quantum statistical problem for a non-commutative gas within a volume $V$ for the purpose of studying its thermodynamics investigating how the introduction of the non-commutativity changes the system. We shall focus our studies in obtaining relevant thermodynamic quantities such as pressure, internal energy, particle number, specific heat, and the Bose-Einstein condensate in the ultra-relativistic (UR) and non-relativistic (NR) limits.

\subsection {Thermodynamics of massive bosons gas}
\label{BLMa}

We will first consider a situation that the system has no antibosons. In grand canonical ensemble we use the Hamiltonian defined in Eq.~(\ref{hamil2}) written in terms of the momentum  $k$ 
\begin{eqnarray}
\Xi &=&{\rm Tr}\,e^{-\beta(H_{NC}-\mu N)}, \nonumber\\
&=& \prod_{\vec{k}}\sum_{m_{\vec{k}},n_{\vec{k}}=0}^{\infty} e^{-\beta(\omega_{\vec{k}}\Lambda_{\vec{k}}^{1}-\mu)n_{\vec{k}}} e^{-\beta(\omega_{\vec{k}}\Lambda_{\vec{k}}^{2}-\mu)m_{\vec{k}}}, \nonumber\\
&=&  \prod_{\vec{k}}\left(\frac{1}{1-ze^{-\beta \omega_{\vec{k}} \Lambda_{\vec{k}}^{1}}}\right) \left(\frac{1}{1-ze^{-\beta \omega_{\vec{k}} \Lambda_{\vec{k}}^{2}}}\right),
\end{eqnarray}
where $N=\sum_{\vec{k}}(N_{\vec{k}}^{1}+N_{\vec{k}}^{2})$ is the operator for the total number of bosons and we are using the system of units $k_{B}=\hbar=c=1$;\, $z=e^{\beta\mu}$ is the fugacity, $\mu$ is the chemical potential and $\beta=1/T$.

Thus, the natural logarithm of the grand partition function is
\begin{eqnarray}
\label{ln1a}
\ln\Xi = - \sum_{\vec{k}}\left[ \ln(1-ze^{-\beta \omega_{\vec{k}} \Lambda_{\vec{k}}^{1}})+ \ln(1-ze^{-\beta \omega_{\vec{k}} \Lambda_{\vec{k}}^{2}})\right].
\end{eqnarray}
We are dealing with a gas formed by $N$ harmonic oscillators within a volume
 $V \sim R^3$. In the thermodynamic limit $V \rightarrow \infty$ $(R \rightarrow \infty)$ the sum is replaced by an integral in the momenta
\begin{equation}
\sum_{\vec{k}} \rightarrow V\int{\frac{d^3k}{(2\pi)^3}} \rightarrow \frac{V}{2\pi^{2}} \int dk k^2. \nonumber
\end{equation}
This makes the Eq.~(\ref{ln1a}) given by
\begin{eqnarray}
\label{ln2a}
\ln\Xi(\beta,V,z) =-\frac{V}{2\pi^{2}} \int_{0}^{\infty} dk k^2\left[ \ln(1-z\mbox{e}^{-\beta E^1(k)})+  \ln(1-z\mbox{e}^{-\beta E^2(k)})\right],
\end{eqnarray}
where \footnote{We shall adopt $g=1/4\pi$.},
\begin{subeqnarray}\label{modener}
\slabel{mdene1}
E^{1}(k)&=&\omega\sqrt{1+\left(\frac{\omega \theta(k)}{8\pi}\right)^2} - \frac{\omega^2\theta(k)}{8\pi} \\
\slabel{mdenerx}
E^{2}(k)&=&\omega\sqrt{1+\left(\frac{\omega\theta(k)}{8\pi}\right)^2} + \frac{\omega^2\theta(k)}{8\pi} 
\end{subeqnarray}
with
\begin{equation}
\omega=\sqrt{k^2+m^2} \,\,\,\,\,\,\,\,\,\,\, \mbox{and} \,\,\,\,\,\,\,\,\,\,\, \theta(k)=\theta \mbox{e}^{-\frac{1}{2}\sigma^2 k^2}.
\end{equation}
Given the complexity of the quantum statistical problem  here formulated, it is not possible to find exact analytic solutions for Eq.~(\ref{ln2a}), because $E^{i}(k)$ is a non trivial function of $k$, $\theta$ and $\sigma$. An alternative to overcome these difficulties is to find suitable approximations able to provide solutions that enable the advancement of our investigations with analytic calculations of the system properties under study. This procedure is made through ultra-relativistic and non-relativistic limits. Alternatively we also treat the problem exactly using numerical approaches.

In this paper we obtain the non-commutative bosonic gas properties developing both procedures, numerical and approximate, in view of checking validation limits of the UR and NR approximation. The relevant quantities for this investigation are the total number of bosons $N$, the internal energy $U$ and the pressure $p$ given by the equations
\begin{subeqnarray} \label{statea}  
\slabel{numeroa}
N&=&z\frac{\partial}{\partial z}\ln\Xi(\beta,V,z),\\
\slabel{energiaa}
U&=&-\frac{\partial}{\partial\beta}\ln\Xi(\beta,V,z),\\
\slabel{pressaoa}
p&=&\frac{1}{\beta V}\ln\Xi(\beta,V,z).
\end{subeqnarray}
We can also analyze the specific heat at constant volume defined by
\begin{eqnarray}
\label{Cv1a}
c_{V}=\frac{1}{N}\left(\frac{\partial U}{\partial T}\right)_{V}=\frac{-\beta^2}{N}\left(\frac{\partial U}{\partial \beta}\right)_{V},
\end{eqnarray}
where the partial derivative $\partial U/\partial \beta$ can be calculated through
\begin{eqnarray}
\label{Cv2a}
\left(\frac{\partial U}{\partial \beta}\right)_{V}&=&\left(\frac{\partial U}{\partial \beta}\right)_{z,V} + \left(\frac{\partial U }{\partial z}\right)_{\beta,V}\left(\frac{\partial z}{\partial \beta}\right)_{N,V} \nonumber \\ 
&=&\left(\frac{\partial U }{\partial \beta}\right)_{z,V} - \left(\frac{\partial U }{\partial z}\right)_{\beta,V}\left(\frac{\partial N}{\partial \beta}\right)_{z,V}\left/  \left(\frac{\partial N}{\partial z}\right)_{\beta,V}\right. .
\end{eqnarray}
Replacing the equation (\ref{ln2a}) into (\ref{statea}) we find the expressions
\begin{subeqnarray}\label{ssss}
\slabel{numero2a}
N&=&\frac{V}{2\pi^2}\int{dk k^2 \left\{ \frac{1}{z^{-1}\mbox{e}^{\beta E^1(k)} -1}+ \frac{1}{z^{-1}\mbox{e}^{\beta E^2(k)} -1}\right\}},  \\ \nonumber \\
\slabel{energia2a}
U&=& \frac{V}{2\pi^2}\int{dk k^2 \left\{ \frac{E^1(k)}{z^{-1}\mbox{e}^{\beta E^1(k)} -1}+ \frac{E^2(k)}{z^{-1}\mbox{e}^{\beta E^2(k)} -1}\right\}}, \\ \nonumber\\
\slabel{pressao2a}
p&=&-\frac{1}{2\pi^{2}\beta} \int dk k^2\left\{ \ln(1-z\mbox{e}^{-\beta E^1(k)})+  \ln(1-z\mbox{e}^{-\beta E^2(k)})\right\}.
\end{subeqnarray}
Here, the total number of bosons $N=N_1+N_2$ is a conserved quantity and an important realization that must be done is to make $N^1$ and $N^2$ positive definite, so that we find the transition temperature $T_0$ at which a relatively high fraction of `atoms' begins to condensate on the lower energy state by taking 
\begin{eqnarray}
\label{mu_ca}
\mu = m' = m\sqrt{1+\left(\frac{m\theta}{8\pi}\right)^2}-\frac{m^2\theta}{8\pi}.  
\end{eqnarray} 
For temperatures greater then $T_0$ we must have $\mu<m'$ and the system becomes  completely excited.
Starting from the equations (\ref{ssss}) we calculate the specific heat via partial derivative  $\partial U/\partial \beta$ given by Eq.~(\ref{Cv2a}) for $T>T_0$ by using
\begin{subeqnarray} \label{diffsa}
\slabel{diffSM1a}
\left(\frac{\partial U }{\partial \beta}\right)_{z,V}&=& -\frac{V}{2\pi^2} \int_{0}^{\infty}{dk k^2 \left\{ \left(\frac{E^1(k)}{z^{-1}\mbox{e}^{\beta E^{1}(k)}-1}\right)^2 z^{-1}\mbox{e}^{\beta E^{1}(k)} + \left(\frac{E^2(k)}{z^{-1}\mbox{e}^{\beta E^{2}(k)}-1}\right)^2 z^{-1}\mbox{e}^{\beta E^{2}(k)} \right\}},  \nonumber \\ \nonumber \\
\slabel{diffSM2a}
\left(\frac{\partial U }{\partial z}\right)_{\beta,V}&=& -\frac{1}{z}\left(\frac{\partial N}{\partial \beta}\right)_{z,V}=  \frac{V z^{-1}}{2\pi^2} \int_{0}^{\infty}{dk k^2 \left\{\frac{E^1(k) z^{-1} \mbox{e}^{\beta E^{1}(k)}}{(z^{-1}\mbox{e}^{\beta E^{1}(k)}-1)^2}  + \frac{E^2(k) z^{-1} \mbox{e}^{\beta E^{2}(k)}}{(z^{-1}\mbox{e}^{\beta E^{2}(k)}-1)^2}  \right\}},  \nonumber \\ \nonumber \\  
\slabel{diffSM3a}
 \left(\frac{\partial N}{\partial z}\right)_{\beta,V}&=& \frac{V z^{-1}}{2\pi^2} \int_{0}^{\infty}{dk k^2 \left\{\frac{z^{-1}\mbox{e}^{\beta E^{1}(k)}}{(z^{-1}\mbox{e}^{\beta E^{1}(k)}-1)^2}  + \frac{z^{-1}\mbox{e}^{\beta E^{2}(k)}}{(z^{-1}\mbox{e}^{\beta E^{2}(k)}-1)^2}  \right\}}.  \nonumber \\ \nonumber
\end{subeqnarray}
For $T<T_0$ we use 
\begin{subeqnarray}
\slabel{difgdsa}
\left(\frac{\partial U }{\partial \beta}\right)_{V}&=& -\frac{V}{2\pi^2} \int_{0}^{\infty}{dk k^2 \left\{ \frac{E^1(k) \left(E^1(k)-m'\right) z^{-1} \mbox{e}^{\beta E^{1}(k)}}{\left(z^{-1} \mbox{e}^{\beta E^{1}(k)}-1\right)^2} + \frac{E^2(k) \left(E^2(k)-m'\right) z^{-1}\mbox{e}^{\beta E^{2}(k)}}{\left(z^{-1}\mbox{e}^{\beta E^{2}(k)}-1 \right)^2}  \right\}}, \nonumber \\ \nonumber
\end{subeqnarray}
 where $z =e^{\beta m'}$.

\subsubsection{The ultra-relativistic limit} \label{secA}

The approximate calculations are made from the expression (\ref{ln2a}) by expanding the integrand as a power series of the fugacity $z$  
\begin{eqnarray}
\label{ln3a}
\ln\Xi= \frac{V}{2\pi^2} \int dk k^{2}\left\{z\left( e^{-\beta E^{1}(k)} + e^{-\beta E^{2}(k)}\right)
+\frac{1}{2} z^2 \left( e^{-2\beta E^{1}(k)} + e^{-2\beta E^{2}(k)}\right) +...\right\}. 
\end{eqnarray}
By using this approach, the ultra-relativistic limit is computed by making $m \ll T_0$ $(\beta_0 m \ll 1)$, so we expand the equation above up to first order in $\beta m$ and then we expand in the parameters $\theta$ and $\sigma$.  Assuming $\theta/\beta < 1$ and  $\sigma/\beta < 1$ we find as result that the main terms to this approximation  are
{\small
\begin{eqnarray} 
\label{lnaprMa}
\ln\Xi (\beta,V,z) &=& \frac{V}{2\pi^2} \left\{\frac{4}{\beta^3}g_{4}(z) +\frac{75\theta^2}{8\beta^5\pi^2}g_{6}(z) -\frac{2205\theta^2\sigma^2}{4\beta^7\pi^2}g_{8}(z) +\frac{33075\theta^4}{1024 \beta^7\pi^4}g_{8}(z)-\frac{654885\theta^4\sigma^2}{64 \beta^9\pi^4}g_{10}(z)+\frac{2837835\theta^6}{16384 \beta^9\pi^6}g_{10}(z)\right\}. \nonumber \\
\end{eqnarray}}
Then from the partition function we find the equations below at ultra-relativistic limit
{\small
\begin{subeqnarray} 
\slabel{numprMa}
N &=& \frac{V}{2\pi^2}\left\{\frac{4}{\beta^3}g_{3}(z) +\frac{75\theta^2}{8\beta^5\pi^2}g_{5}(z) -\frac{2205\theta^2\sigma^2}{4\beta^7\pi^2}g_{7}(z) +\frac{33075\theta^4}{1024 \beta^7\pi^4}g_{7}(z)-\frac{654885\theta^4\sigma^2}{64 \beta^9\pi^4}g_{9}(z)+\frac{2837835\theta^6}{16384\beta^9\pi^6}g_{9}(z)\right\}, \nonumber \\ \\
\slabel{eneprMa}
U &=& \frac{V}{2\pi^2}\left\{ \frac{12}{\beta^4}g_{4}(z) +\frac{375\theta^2}{8\beta^6\pi^2}g_{6}(z) -\frac{15435\theta^2\sigma^2}{4\beta^8\pi^2}g_{8}(z) +\frac{231525\theta^4}{1024 \beta^8\pi^4}g_{8}(z)-\frac{5893965\theta^4\sigma^2}{64 \beta^{10}\pi^4}g_{10}(z)+\frac{25540515\theta^6}{16384\beta^{10}\pi^6}g_{10}(z)\right\} \nonumber \\ \\ 
\slabel{pressMa}
p &=& \frac{1}{2\pi^2}\left\{\frac{4}{\beta^4}g_{4}(z) +\frac{75\theta^2}{8\beta^6\pi^2}g_{6}(z) -\frac{2205\theta^2\sigma^2}{4\beta^8\pi^2}g_{8}(z) +\frac{33075\theta^4}{1024 \beta^8\pi^4}g_{8}(z)-\frac{654885\theta^4\sigma^2}{64 \beta^{10}\pi^4}g_{10}(z)+\frac{2837835\theta^6}{16384 \beta^{10}\pi^6}g_{10}(z)\right\}  \nonumber \\
\end{subeqnarray}}
where  $g_{\alpha}(z)$ is the Bose function defined as
\begin{eqnarray}
g_{\alpha}(z)=\sum_{n=1}^{\infty} \frac{z^n}{n^\alpha}. \nonumber
\end{eqnarray}
This function diverges at $z = 1$ when $\alpha \leq 1$, but becomes the Riemann zeta function $\zeta (\alpha)$ when $\alpha>1$.

The specific heat (\ref{Cv1a}) is computed using the equations  (\ref{numprMa}) and (\ref{eneprMa}). As we previously stated for $T<T_0$ we have $z =e^{\beta m'}$ then
{\small
\begin{subeqnarray} 
\left(\frac{\partial U }{\partial \beta}\right)_{V}&=&\frac{V}{2\pi^2}\left\{ \frac{12}{\beta^4}\left(m'g_{3}(z)-\frac{4}{\beta}g_{4}(z)\right) + \frac{375\theta^2}{8\beta^6\pi^2}\left( m'g_{5}(z)-\frac{6}{\beta}g_{6}(z)\right)+\left(\frac{231525\theta^4}{1024\beta^8\pi^4}-\frac{15435\theta^2\sigma^2}{4\beta^8\pi^2}\right)\left(m'g_{7}(z)-\frac{8}{\beta}g_{8}(z)\right) \right.\nonumber \\
& +&\left. \left(\frac{25540515\theta^6}{16384\beta^{10}\pi^6}-\frac{5893965\theta^4\sigma^2}{64\beta^{10}\pi^4}\right)\left(m'g_{9}(z)-\frac{10}{\beta}g_{10}(z)\right)\right\}.  \nonumber \\ \nonumber
\end{subeqnarray}}
For $T>T_0$ we use the following equations
{\small
\begin{subeqnarray}
\left(\frac{\partial U }{\partial \beta}\right)_{z,V} &=& -\frac{V}{2\pi^2}\left\{\frac{48}{\beta^5}g_{4}(z) +\frac{1125\theta^2}{4\beta^7\pi^2}g_{6}(z)- \frac{30870\theta^2\sigma^2}{\beta^9\pi^2}g_{8}(z)+\frac{231525\theta^4}{128\beta^9\pi^4}g_{8}(z)-\frac{29469825\theta^4\sigma^2}{32\beta^{11}\pi^4}g_{10}(z)+\frac{127702575\theta^6}{8192\beta^{11}\pi^6}g_{10}(z) \right\}, \nonumber \\ \nonumber \\
\left(\frac{\partial U }{\partial z}\right)_{\beta,V}&=& \frac{Vz^{-1}}{2\pi^2}\left\{  \frac{12}{\beta^4}g_{3}(z) +\frac{375\theta^2}{8\beta^6\pi^2}g_{5}(z) -\frac{15435\theta^2\sigma^2}{4\beta^8\pi^2}g_{7}(z) +\frac{231525\theta^4}{1024 \beta^8\pi^4}g_{7}(z)-\frac{5893965\theta^4\sigma^2}{64 \beta^{10}\pi^4}g_{9}(z)+\frac{25540515\theta^6}{16384\beta^{10}\pi^6}g_{9}(z)\right\}, \nonumber\\ \nonumber \\
\left(\frac{\partial N}{\partial z}\right)_{\beta,V} &=& \frac{Vz^{-1}}{2\pi^2}\left\{\frac{4}{\beta^3}g_{2}(z) +\frac{75\theta^2}{8\beta^5\pi^2}g_{4}(z) -\frac{2205\theta^2\sigma^2}{4\beta^7\pi^2}g_{6}(z) +\frac{33075\theta^4}{1024 \beta^7\pi^4}g_{6}(z) - \frac{654885\theta^4\sigma^2}{64 \beta^9\pi^4}g_{8}(z)+  \frac{2837835\theta^6}{16384 \beta^9\pi^6}g_{8}(z)\right\}.  \nonumber \\\nonumber
\end{subeqnarray}}

\subsubsection{Non-relativistic limit}

To take the non-relativistic limit $m \gg T_0$ and we can expand the modified energies (\ref{modener}) in powers of $k^2/m^2$ up to first order and with respect to the parameters $\theta$ and $\sigma$. We obtain the following results as the major contribution to NR limit
\begin{subeqnarray} \label{enemodexp}
E^1(k)&\approx& m+ \frac{k^2}{2m} -\frac{m^2\theta}{8\pi} \\ 
E^2(k)&\approx&m+ \frac{k^2}{2m} +\frac{m^2\theta}{8\pi}.
\end{subeqnarray}
Now we substitute the energies above into Eq.~(\ref{ln3a}), such that we find
\begin{eqnarray}
\label{lnNRa}
\ln\Xi=V\left(\frac{m}{2\pi\beta}\right)^{3/2}\left[g_{5/2}\left(ze^{\beta\left(\frac{m^2\theta}{8\pi}-m\right)}\right)+g_{5/2}\left(ze^{-\beta\left(\frac{m^2\theta}{8\pi}+m\right)}\right)\right].
\end{eqnarray}
Given the partition function we find the expressions for the total number of particles and internal energy
\begin{eqnarray}
\label{numNR}
N&=&V\left(\frac{m}{2\pi\beta}\right)^{3/2}\left[g_{3/2}\left(ze^{\beta\left(\frac{m^2\theta}{8\pi}-m\right)}\right)+g_{3/2}\left(ze^{-\beta\left(\frac{m^2\theta}{8\pi}+m\right)}\right)\right], \nonumber \\ \\
\label{eneNR}
U&=&N_1\left(m-\frac{m^2\theta}{8\pi}\right) + N_2\left(m+\frac{m^2\theta}{8\pi}\right)+\frac{3V}{2\beta}\left(\frac{m}{2\pi\beta}\right)^{3/2} \left[g_{5/2}\left(z e^{\beta\left(\frac{m^2\theta}{8\pi}-m\right)}\right) + g_{5/2}\left(z e^{-\beta\left(\frac{m^2\theta}{8\pi}+m\right)}\right)\right], \nonumber \\
\end{eqnarray}
where we see the non-commutative terms acting  differently in the rest mass of the two types of bosons. In this approximation, the critical chemical potential and the critical temperature  are given by
\begin{subeqnarray}
\mu_{0}&=&m-\frac{m^2\theta}{8\pi} \,\,\,\,\,\,\,\, \mbox{and} \,\,\,\,\,\,\,\, N = V\left(\frac{m}{2\pi\beta_0}\right)^{3/2}\left[g_{3/2}\left(1\right)+g_{3/2}\left(e^{-\beta_0\left(\frac{m^2\theta}{4\pi}\right)}\right)\right]. \\ \nonumber
\end{subeqnarray} 
We can see that the non-commutative theory  modifies the thermodynamics of the gas introducing corrections that does not exist in the standard BEC. All the previous non-relativistic results known in the literature are recovered as $\theta \rightarrow 0$.

\subsection {Thermodynamics of massive bosons-antibosons gas}
\label{BLSM}

To describe the system with antibosons we express  $\Xi={\rm Tr}\,e^{-\beta(H_{NC}-\mu Q)}$ where $Q=\sum_{\vec{k}}(N_{\vec{k}}^{1}-N_{\vec{k}}^{2})$ is a operator which corresponds to a conserved quantum number generically referred as to charge. The grand partition function is now explicitly given by
\begin{eqnarray}
\Xi &=& \prod_{\vec{k}}\sum_{m_{\vec{k}},n_{\vec{k}}=0}^{\infty} e^{-\beta(\omega_{\vec{k}}\Lambda_{\vec{k}}^{1}-\mu)n_{\vec{k}}} e^{-\beta(\omega_{\vec{k}}\Lambda_{\vec{k}}^{2}+\mu)m_{\vec{k}}} \nonumber\\
&=&  \prod_{\vec{k}}\left(\frac{1}{1-ze^{-\beta \omega_{\vec{k}} \Lambda_{\vec{k}}^{1}}}\right) \left(\frac{1}{1-z^{-1}e^{-\beta \omega_{\vec{k}} \Lambda_{\vec{k}}^{2}}}\right).
\end{eqnarray}
Recall that we have previously identified $\Lambda_{\vec{k}}^{1}$ and $\Lambda_{\vec{k}}^{2}$ in (\ref{eqs-45-energ}) as the energies of the particle and antiparticle, as well discussed in \cite{Camona2}.
The grand partition function in the thermodynamic limit is
\begin{eqnarray}
\label{ln2}
\ln\Xi(\beta,V,z) =-\frac{V}{2\pi^{2}} \int_{0}^{\infty} dk k^2\left[ \ln(1-z\mbox{e}^{-\beta E^1(k)})+  \ln(1-z^{-1}\mbox{e}^{-\beta E^2(k)})\right].
\end{eqnarray}
Analogously to the previous section, we obtain the expressions for the net charge $Q$, the internal energy $U$ and the pressure $p$ through the equations
\begin{subeqnarray} \label{state}  
\slabel{numero}
Q&=&z\frac{\partial}{\partial z}\ln\Xi(\beta,V,z),\\
\slabel{energia}
U&=&-\frac{\partial}{\partial\beta}\ln\Xi(\beta,V,z),\\
\slabel{pressao}
p&=&\frac{1}{\beta V}\ln\Xi(\beta,V,z),
\end{subeqnarray}
and the specific heat at constant volume is
\begin{eqnarray}
\label{Cv1}
c_{V}=\frac{-\beta^2}{Q}\left(\frac{\partial U }{\partial \beta}\right)_{V}.
\end{eqnarray}
Replacing the equation (\ref{ln2}) into (\ref{numero}) we find the expression for the charge
\begin{eqnarray}
\label{numero2}
Q=\frac{V}{2\pi^2}\int_{0}^{\infty}{dk k^2 \left\{ \frac{1}{z^{-1}\mbox{e}^{\beta E^1(k)} -1}- \frac{1}{z\mbox{e}^{\beta E^2(k)} -1}\right\}}.
\end{eqnarray}
In this context, we have a conservation of the number $Q=N_1-N_2$, where $N_1$ and $N_2$ are the total number of bosons and antibosons, respectively. Making $N^1$ and $N^2$ positive definite we get to the conclusion 
\begin{eqnarray}
\label{q2}
-\omega \sqrt{1+\left(\frac{\omega\theta(k)}{8\pi}\right)^2}-\frac{\omega^2\theta(k)}{8\pi} \leq \mu \leq \omega\sqrt{1+\left(\frac{\omega\theta(k)}{8\pi}\right)^2}-\frac{\omega^2\theta(k)}{8\pi},
\end{eqnarray}
where we suppose that the bosons number exceeds the antibosons number, i.e., $N_1>N_2$  and thus $Q>0$, then we  find $\mu>-\omega^2\theta (k)/8\pi$ and the transition temperature $T_0$ is obtained by taking 
\begin{eqnarray}
\label{mu_c}
\mu = m' = m\sqrt{1+\left(\frac{m\theta}{8\pi}\right)^2}-\frac{m^2\theta}{8\pi}.  
\end{eqnarray}
Note that the Eq.(\ref{numero2}) is actually the number of excited charges, $Q-Q_0$, where $Q_0$ is the number of charge in the ground state. So at $\beta=\beta_0$ and $\mu=m'$ we can get total charge density $\rho=Q/V$ written as
\begin{eqnarray}
\label{rhop}
\rho=\frac{1}{2\pi^2}\int_{0}^{\infty}{dk k^2  \frac{\sinh\left[\beta_0 \left(\frac{\omega^2\theta(k)}{8\pi}+m' \right)\right]}{\cosh\left[\beta_0\omega\sqrt{1+\left(\frac{\omega\theta(k)}{8\pi}\right)^2}\right]-\cosh\left[\beta_0 \left(\frac{\omega^2\theta(k)}{8\pi}+m' \right)\right]}}.
\end{eqnarray}
Using Eq.(\ref{energia}) we get the internal energy of the charged system  
\begin{eqnarray}
\label{energia2}
U&=& \frac{V}{2\pi^2}\int_{0}^{\infty}{dk k^2 \left\{ \frac{E^1(k)}{z^{-1}\mbox{e}^{\beta E^1(k)} -1}+ \frac{E^2(k)}{z\mbox{e}^{\beta E^2(k)} -1}\right\}},
\end{eqnarray}
which can also be written as
\begin{eqnarray}
\label{eenee}
U=\frac{V}{2\pi^2}\int_{0}^{\infty}{dk k^2\omega \frac{\sqrt{1+\left(\frac{\omega\theta(k)}{8\pi}\right)^2} \left[\cosh\left[\beta \left(\frac{\omega^2\theta(k)}{8\pi}+\mu \right)\right] -e^{-\beta\omega\sqrt{1+\left(\frac{\omega\theta(k)}{8\pi}\right)^2} }\right] - \frac{\omega\theta(k)}{8\pi}\sinh\left[\beta \left(\frac{\omega^2\theta(k)}{8\pi}+\mu \right)\right] }{\cosh\left[\beta\omega\sqrt{1+\left(\frac{\omega\theta(k)}{8\pi}\right)^2}\right]-\cosh\left[\beta\left(\frac{\omega^2\theta(k)}{8\pi}+\mu\right)\right]}}, \nonumber \\
\end{eqnarray}
and the pressure is
\begin{eqnarray}
\label{pressao2}
p =-\frac{1}{2\pi^{2}\beta} \int_{0}^{\infty} dk k^2\left[ \ln(1-z\mbox{e}^{-\beta E^1(k)})+  \ln(1-z^{-1}\mbox{e}^{-\beta E^2(k)})\right].
\end{eqnarray}
Starting from the equations (\ref{numero2}) and (\ref{energia2}) we can calculate the specific heat.  For $T>T_0$
\begin{subeqnarray} \label{diffs}
\slabel{diffSM1}
\left(\frac{\partial U }{\partial \beta}\right)_{z,V}&=& -\frac{V}{2\pi^2} \int_{0}^{\infty}{dk k^2 \left\{ \left(\frac{E^1(k)}{z^{-1}\mbox{e}^{\beta E^{1}(k)}-1}\right)^2 z^{-1}\mbox{e}^{\beta E^{1}(k)} + \left(\frac{E^2(k)}{z\mbox{e}^{\beta E^{2}(k)}-1}\right)^2 z\mbox{e}^{\beta E^{2}(k)} \right\}},  \nonumber \\  \\
\slabel{diffSM2}
\left(\frac{\partial U }{\partial z}\right)_{\beta,V}&=& -\frac{1}{z}\left(\frac{\partial Q}{\partial \beta}\right)_{z,V}=  \frac{V z^{-1}}{2\pi^2} \int_{0}^{\infty}{dk k^2 \left\{\frac{E^1(k) z^{-1} \mbox{e}^{\beta E^{1}(k)}}{(z^{-1}\mbox{e}^{\beta E^{1}(k)}-1)^2}  - \frac{E^2(k) z \mbox{e}^{\beta E^{2}(k)}}{(z\mbox{e}^{\beta E^{2}(k)}-1)^2}  \right\}},  \nonumber \\  \\  
\slabel{diffSM3}
 \left(\frac{\partial Q}{\partial z}\right)_{\beta,V}&=& \frac{V z^{-1}}{2\pi^2} \int_{0}^{\infty}{dk k^2 \left\{\frac{z^{-1}\mbox{e}^{\beta E^{1}(k)}}{(z^{-1}\mbox{e}^{\beta E^{1}(k)}-1)^2}  - \frac{z\mbox{e}^{\beta E^{2}(k)}}{(z\mbox{e}^{\beta E^{2}(k)}-1)^2}  \right\}},  \nonumber \\ 
\end{subeqnarray}
and for $T<T_0$
\begin{eqnarray}
\left(\frac{\partial U }{\partial \beta}\right)_{V}&=& -\frac{V}{2\pi^2} \int_{0}^{\infty}{dk k^2 \left\{ \frac{E^1(k) \left(E^1(k)-m'\right) z^{-1} \mbox{e}^{\beta E^{1}(k)}}{\left(z^{-1} \mbox{e}^{\beta E^{1}(k)}-1\right)^2} + \frac{E^2(k) \left(E^2(k)+m'\right) z\mbox{e}^{\beta E^{2}(k)}}{\left(z\mbox{e}^{\beta E^{2}(k)}-1 \right)^2}  \right\}}, \nonumber
\end{eqnarray}
where we have $z =e^{\beta m'}$.

\subsubsection{The ultra-relativistic limit}

The approximate calculations are obtained from the expression (\ref{ln2}) by expanding the integrand as a power series of the fugacity $z_1=z$ and $z_2=z^{-1}$ 

\begin{eqnarray}
\label{ln3}
\ln\Xi= \frac{V}{2\pi^2} \int_{0}^{\infty} dk k^{2}\left\{\left(e^{-\beta E^1(k)}z_{1} + \frac{1}{2} e^{-2\beta E^1(k)} z_{1}^{2} + ...\right)+\left(e^{-\beta E^2(k)}z_{2} + \frac{1}{2} e^{-2\beta E^2(k)} z_{2}^{2} + ...\right)\right\}. \nonumber \\
\end{eqnarray}
This is analogous to section \ref{secA}, we take the ultra-relativistic limit and then we find 
\begin{eqnarray} 
\label{ln4}
\ln\Xi(\beta,V,z) &=& \frac{V}{2\pi^2} \left\{\frac{2}{\beta^3} \left[g_{4}(z)+ g_{4}(z^{-1})\right]
+\frac{3\theta}{\beta^4\pi}\left[g_{5}(z) - g_{5}(z^{-1})\right] -\frac{45\theta\sigma^2}{\beta^6\pi}\left[g_{7}(z) - g_{7}(z^{-1})\right] \right. \nonumber \\ 
&+& \left. \frac{75\theta^2}{16\beta^5\pi^2}\left[g_{6}(z) + g_{6}(z^{-1})\right] + \frac{525\theta^3}{64\beta^6\pi^3}\left[g_{7}(z) - g_{7}(z^{-1})\right]   
 \right\}. \nonumber \\ 
\end{eqnarray}
Again, since the partition function is obtained we can determine the excited charge density $\rho_e$,  internal energy density $u=U/V$ and pressure $p$
\begin{subeqnarray}\label{statesapprox}
\slabel{numero1}
\rho_e&=&\frac{1}{\pi^2\beta^3} \left[g_{3}(z)- g_{3}(z^{-1})\right]
+\frac{3\theta}{2\beta^4\pi^3}\left[g_{4}(z) + g_{4}(z^{-1})\right] -\frac{45\theta\sigma^2}{2\beta^6\pi^3}\left[g_{6}(z) + g_{6}(z^{-1})\right] \nonumber \\ 
&+&  \frac{75\theta^2}{32\beta^5\pi^4}\left[g_{5}(z) - g_{5}(z^{-1})\right] + \frac{525\theta^3}{128\beta^6\pi^5}\left[g_{6}(z) + g_{6}(z^{-1})\right], \nonumber \\ \\ 
\slabel{energia1}
u &=& \frac{3}{\pi^2\beta^4} \left[g_{4}(z)+ g_{4}(z^{-1})\right]
+\frac{6\theta}{\beta^5\pi^3}\left[g_{5}(z) - g_{5}(z^{-1})\right] -\frac{135\theta\sigma^2}{\beta^7\pi^3}\left[g_{7}(z) - g_{7}(z^{-1})\right] \nonumber \\ 
&+&  \frac{375\theta^2}{32\beta^6\pi^4}\left[g_{6}(z) + g_{6}(z^{-1})\right] + \frac{1575\theta^3}{64\beta^7\pi^5}\left[g_{7}(z) - g_{7}(z^{-1})\right], \nonumber \\ \\
\slabel{pressao1}
p &=&  \frac{1}{\pi^2\beta^4} \left[g_{4}(z)+ g_{4}(z^{-1})\right]
+\frac{3\theta}{2\beta^5\pi^3}\left[g_{5}(z) - g_{5}(z^{-1})\right] -\frac{45\theta\sigma^2}{2\beta^7\pi^3}\left[g_{7}(z) - g_{7}(z^{-1})\right] \nonumber \\ 
&+& \frac{75\theta^2}{32\beta^6\pi^4}\left[g_{6}(z) + g_{6}(z^{-1})\right] + \frac{525\theta^3}{128\beta^7\pi^5}\left[g_{7}(z) - g_{7}(z^{-1})\right]. \nonumber \\ 
\end{subeqnarray}
But in the ultra-relativistic limit $\beta m \ll 1 \Rightarrow \beta \mu \ll 1$ where $\mu \leq m'$, then 
\begin{eqnarray}
e^{\pm\beta\mu}\approx 1\pm \beta\mu \,\,\,\,\,\,\,\, \Rightarrow \,\,\,\,\,\,\,\, g_{\alpha}\left(e^{\pm\beta\mu}\right) \approx g_{\alpha}(1) \pm \beta\mu g_{\alpha-1}(1).
\end{eqnarray}
This approximation leads to the thermodynamic quantities
\begin{subeqnarray}\label{URES}
\slabel{number11}
\rho_e&=& \frac{2 \mu}{\pi^2 \beta^{2}} \zeta(2) + \frac{3\theta}{\beta^{4}\pi^3} \zeta(4)  -\frac{45\theta\sigma^2}{\beta^{6}\pi^3}\zeta(6) + \frac{75 \mu\theta^2}{16\beta^{4}\pi^4} \zeta(4) + \frac{525\theta^3}{64\beta^{6}\pi^5}\zeta(6),  \\ \nonumber \\
\slabel{internal}
u &=& \frac{6}{\pi^2\beta^4}\zeta(4) + \frac{12\mu\theta}{\beta^4\pi^3}\zeta(4) -\frac{270\mu\theta\sigma^2}{\beta^6\pi^3}\zeta(6) + \frac{375\theta^2}{16\beta^6\pi^4}\zeta(6)+ \frac{1575\mu\theta^3}{32\beta^6\pi^5}\zeta(6),  \\ \nonumber \\ 
\slabel{pressure}
p &=&  \frac{2}{\pi^2\beta^4}\zeta(4) + \frac{3\mu\theta}{\beta^4\pi^3}\zeta(4) -\frac{45\mu\theta\sigma^2}{\beta^6\pi^3}\zeta(6) + \frac{75\theta^2}{16\beta^6\pi^4}\zeta(6) + \frac{525\mu\theta^3}{64\beta^6\pi^5}\zeta(6).  
\end{subeqnarray}
So the expression for the total charge density is 
\begin{eqnarray}
\label{numero111}
\rho&=& \frac{2 m'}{\pi^2 \beta_{0}^{2}} \zeta(2) + \frac{3\theta}{\beta_{0}^{4}\pi^3} \zeta(4)  -\frac{45\theta\sigma^2}{\beta_{0}^{6}\pi^3}\zeta(6) + \frac{75 m'\theta^2}{16\beta_{0}^{4}\pi^4} \zeta(4) + \frac{525\theta^3}{64\beta_{0}^{6}\pi^5}\zeta(6).  
\end{eqnarray}
Starting from this equation we can obtain the critical temperature in terms of $\rho$. When $\theta\rightarrow0$ we get $T_0=(3\rho/m)^{1/2}$ which is in agreement with the Haber and Weldon result ~\cite{Harber}. As consequence, for the $m\rightarrow 0$ limit we get that $T_0 \rightarrow \infty$ and hence all net charge of the massless bosons gas resides in the ground state, differently from the non-commutative gas where if $m\rightarrow 0$ we will have $T_0 \sim \left(\pi^3\rho/3\theta\zeta(4)\right)^{1/4}$.

In addition, the specific heat for $T>T_0$ can be found by
\begin{subeqnarray} 
\slabel{aaa}
\left(\frac{\partial U }{\partial \beta}\right)_{z,V}&=& -\frac{V}{\pi^2} \left\{\frac{24}{\beta^5}\zeta(4) + \frac{60\mu\theta}{\beta^5\pi}\zeta(4) -\frac{1890\mu\theta\sigma^2}{\beta^7\pi}\zeta(6) + \frac{1125\theta^2}{8\beta^7\pi^2}\zeta(6)+ \frac{11025\mu\theta^3}{32\beta^7\pi^3}\zeta(6)\right\}, \nonumber \\  \\ 
\slabel{bbb}
\left(\frac{\partial U }{\partial z}\right)_{\beta,V}&=&-\frac{1}{z}\left(\frac{\partial Q}{\partial \beta}\right)_{z,V}=  \frac{V}{z\pi^2}\left\{\frac{6 \mu}{\beta^{3}} \zeta(2) + \frac{12\theta}{\beta^{5}\pi} \zeta(4)  -\frac{270\theta\sigma^2}{\beta^{7}\pi}\zeta(6) + \frac{375 \mu\theta^2}{16\beta^{5}\pi^2} \zeta(4) + \frac{1575\theta^3}{32\beta^{7}\pi^3}\zeta(6)\right\},  \nonumber \\  \\
\slabel{ccc}
 \left(\frac{\partial Q}{\partial z}\right)_{\beta,V}&=& \frac{V}{z\pi^2}\left\{\frac{2}{\beta^{3}} \zeta(2) + \frac{3\mu\theta}{\beta^{3}\pi} \zeta(2)  -\frac{45\mu\theta\sigma^2}{\beta^{5}\pi}\zeta(4) + \frac{75\theta^2}{16\beta^{5}\pi^2} \zeta(4) + \frac{525\mu\theta^3}{64\beta^{5}\pi^3}\zeta(4)\right\},\nonumber \\
\end{subeqnarray}
and for $T<T_0$

\begin{eqnarray}\label{derapro}
\left(\frac{\partial U }{\partial \beta}\right)_{V}&=& -\frac{V}{\pi^2} \left\{\frac{24}{\beta^5}\zeta(4) + \frac{48 m' \theta}{\beta^5\pi}\zeta(4) -\frac{1620 m' \theta\sigma^2}{\beta^7\pi}\zeta(6) + \frac{1125\theta^2}{8\beta^7\pi^2}\zeta(6)+ \frac{4725 m' \theta^3}{16\beta^7\pi^3}\zeta(6)\right\}.\,\,\,\,\,\,\,\,\,\,\,\,\,\,\,\,\,\,\, \nonumber
\end{eqnarray}
\subsubsection{The non-relativistic limit}

In the non-relativistic limit we use the energies (\ref{enemodexp}). We subsequently substitute these energies into Eq.~(\ref{ln3}), to find
\begin{eqnarray}
\label{lnNR}
\ln\Xi=V\left(\frac{m}{2\pi\beta}\right)^{3/2}\left[g_{5/2}\left(e^{\beta\left(\mu+\frac{m^2\theta}{8\pi}-m\right)}\right)+g_{5/2}\left(e^{-\beta\left(\mu+\frac{m^2\theta}{8\pi}+m\right)}\right)\right].
\end{eqnarray}
In the case of non-relativistic limit the temperature is low enough such that $\beta m \gg 1$, then the second term into equation above corresponding to antibosons can be neglected. But still has non-commutative modification in the particle term acting as an additional mass which decreases the boson rest mass. In this approximation the thermodynamic equations are
\begin{subeqnarray}\label{NReee}
\slabel{nuNR}
N&=&V\left(\frac{m}{2\pi\beta}\right)^{3/2}g_{3/2}\left(z e^{\beta\left(\frac{m^2\theta}{8\pi}-m\right)}\right), \\
\slabel{enNR}
U&=&N\left(m-\frac{m^2\theta}{8\pi}\right) + \frac{3V}{2\beta}\left(\frac{m}{2\pi\beta}\right)^{3/2}g_{5/2}\left(z e^{\beta\left(\frac{m^2\theta}{8\pi}-m\right)}\right),\\
\slabel{preNR}
p&=&\frac{1}{\beta}\left(\frac{m}{2\pi\beta}\right)^{3/2}g_{5/2}\left(z e^{\beta\left(\frac{m^2\theta}{8\pi}-m\right)}\right).
\end{subeqnarray}
The specific heat reads
\begin{eqnarray}
c_V= \frac{15}{4}\frac{g_{5/2}(\bar{z})}{g_{3/2}(\bar{z})}-\frac{9}{4}\frac{g_{3/2}(\bar{z})}{g_{1/2}(\bar{z})},
\end{eqnarray}
where we have set $\bar{z}=z e^{\beta\left(\frac{m^2\theta}{8\pi}-m\right)}$. $\bar{z}=1$ for $T\leq T_0$, while $\bar{z}<1$ for $T>T_0$,  we fall back at the  situation of usual non-relativistic Bose gas.

Therefore, in the study of the Bose-Einstein condensation through non-commutative fields arise corrections in the system properties, which can open new investigations such as looking for small deformation effects produced by such systems in high energy physics, astrophysics, cosmology or even in condensed matter. 

\section{Results and discussions} \label{chp:3}

The thermodynamic behaviour of scalar fields was explored in the context of quantum field theory on a non-commutative target space, as  presented in section \ref{chp:2}. This leads to deformation of physical quantities that will be analyzed in detail in this section.  The results are compared to those well-established in the literature by the usual commutative theory. We shall plot graphics of the modified thermodynamic functions in different regions of temperature, which show how the non-commutative parameters $\theta$ and $\sigma$ affect these quantities. These parameters are expected to be related to short distances, thus they cannot be large and were set at the following values $\theta=2.4\times 10^{-14}$ and $\sigma=1.1\times 10^{-16}$. For convenience we assume the value of parameters in arbitrary units.

\subsection{Bose-Einstein condensation without antibosons}

\begin{figure}[!h]
	\centering
		\includegraphics[width=10.0cm,height=6.6cm]{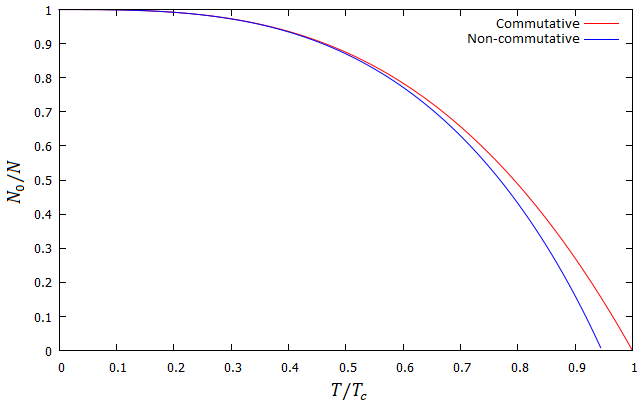}
	\caption[Ratio of particles in the ground state as a function of the temperature]{Comparison of the ratio of particles in the ground state as a function of the temperature for commutative and non-commutative bosons. Fixing $\theta=2.4\times 10^{-14}$, $\sigma=1.1\times 10^{-16}$ and $m=5.1\times 10^{5}$ we choose $n=N/V = 1.0\times 10^{40}$ thus $T_{0} \approx 3.26\times 10^{13}$ and $T_{c} \approx 3.45\times 10^{13}$.\footnote{We will refer $T_c$ and  $T_{0}$ as the critical temperatures of commutative and non-commutative cases, respectively.}}
	\label{fig:num}
\end{figure}

We will present the contributions made on a  system of $N$ fixed massive particles. The critical temperature $T_0$ at which the Bose-Einstein condensation occurs corresponds to $\mu=m'$. For some temperature above $T_0$ we can always find a chemical potential satisfying $\mu<m'$ such that Eq.(\ref{numero2a}) holds. On the other hand, for $T<T_0$ the ground state becomes macroscopically occupied, then the number of  particles in excited states decreases proportionally. This behavior can be seen in the Figure [\ref{fig:num}] for  commutative and non-commutative bosons with the mass adopted as $m=5.1\times 10^{5}$.

\begin{figure}[!h]
	\centering
		\includegraphics[width=10.0cm,height=6.6cm]{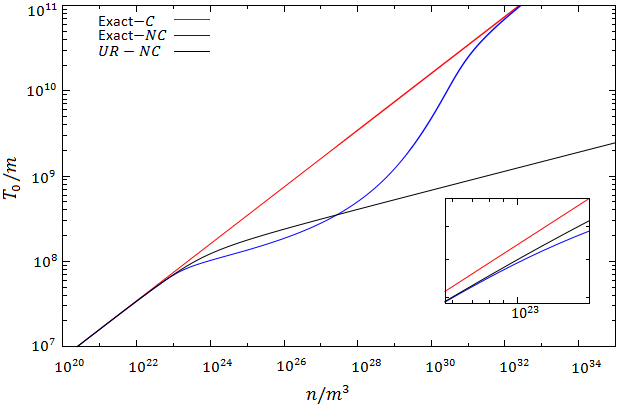}
	 \caption{Results of $T_0$ as a function of density particle $n$ comparing commutative (exact-C), non-commutative (exact-NC) and ultra-relativistic (UR-NC) cases.}
\label{fig:N_T0}
\end{figure}

Analyzing the Figure [\ref{fig:num}] we note that the behavior obtained after introduction of non-commutativity is similar to the commutative case. However, the critical temperature is modified which can also be seen in Figure [\ref{fig:N_T0}], where are shown the numerical results of $T_0$ for different values of particle density, as well as those obtained by the non-commutative ultra-relativistic (UR-NC) regime. For a given particle density there is a temperature region where the transition temperatures of non-commutative curves are lower. This effect begins to appear in the graph in high temperatures $(T\sim 10^{13})$. We can also check the validity of our UR-NC approach that  coincides with the non-commutative exact result in a limited range of temperature. 

\begin{figure}[!h]
	\subfigure[\label{fig:ene}]{\includegraphics[width=9.0cm,height=7.1cm]{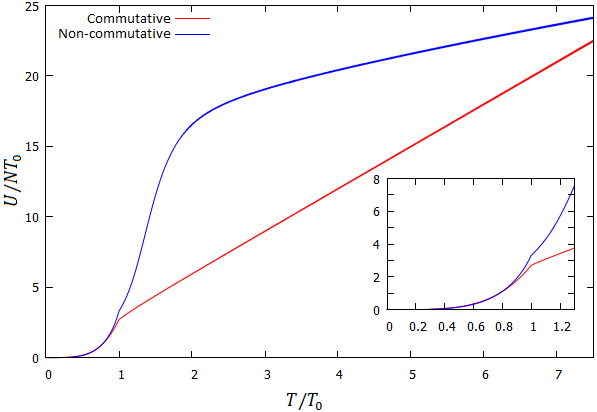}}\\
	\subfigure[\label{fig:cvN}]{\includegraphics[width=9.0cm,height=7.1cm]{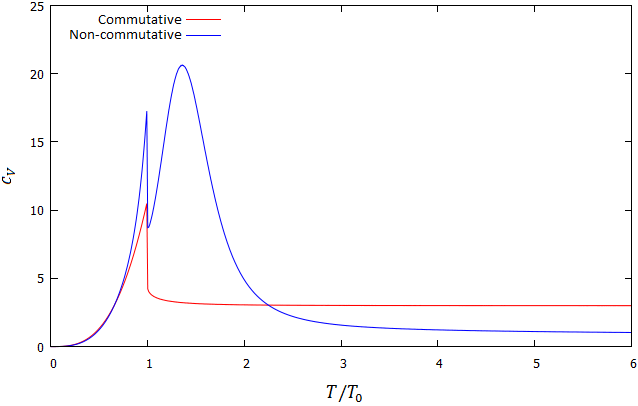}}
 \caption[(a) Internal energy per particle as a function of the temperature. (b) Specific heat as a function of temperature.]{(a)  Internal energy per particle as a function of the temperature. (b) Specific heat as a function of temperature with a fixed number of particles. Using the same parameters of Fig.[\ref{fig:num}].}
\label{fig:enecv}
\end{figure}

Continuing our investigations we study the non-commutative effects on internal energy and specific heat, whose behaviors are shown in Figure [\ref{fig:enecv}]. The results show an increase in the discontinuity at $T=T_0$ of specific heat of the system, that features a second-order phase transition. The gap is equal to $8.55$ for non-commutative bosons while is $6.26$ for commutative bosons. Moreover,  it is interesting to observe that in the region $T>T_0$ arise a lump. A contribution that can be related to the angular momentum term in the non-commutative Hamiltonian. In low temperature physics, these effects can be even more interesting in condensed matter physics, such as in high-Tc superconductivity and thermodynamic anomaly of water. 

\subsection{Bose-Einstein condensation with antibosons}

\begin{figure}[!h]
	\subfigure[\label{fig:char}]{\includegraphics[width=8.5cm,height=6.6cm]{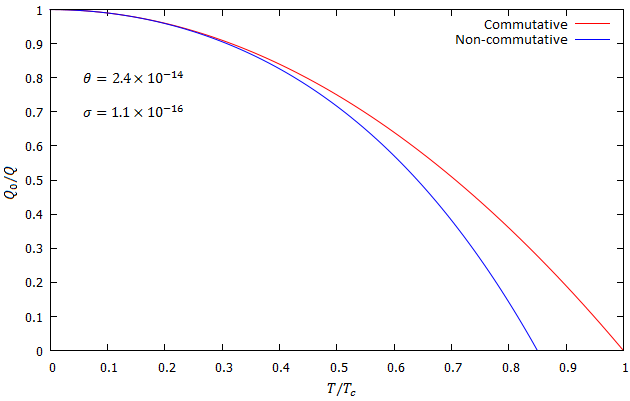}} \\
	\subfigure[\label{fig:chargtheta}]{\includegraphics[width=8.5cm,height=6.6cm]{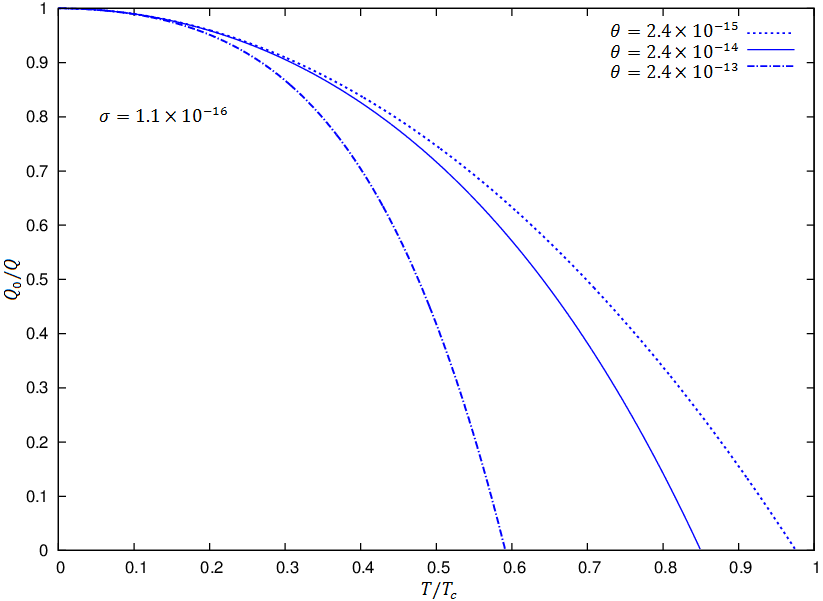}}
	\qquad
	\subfigure[\label{fig:chargsigma}]{\includegraphics[width=7.5cm,height=5.5cm]{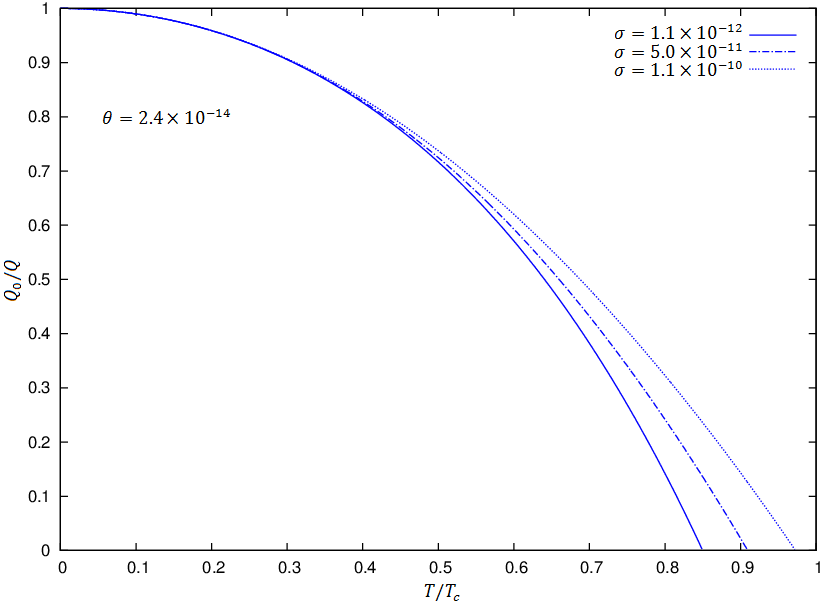}}
\caption[Ratio of charges in the ground state as a function of the temperature.]{Results  for the ratio of charges in the ground state as a function of the temperature choosing $\rho= 6.12\times 10^{24}$ and $m=5.1\times 10^{5}$. (a) Comparing commutative and non-commutative cases, the critical temperatures are $T_{c} \approx 6.0 \times 10^{9}$ for commutative bosons and $T_{0} \approx 5.1 \times 10^{9}$ for non-commutative bosons with $\theta=2.4\times 10^{-14}$ and $\sigma=1.1\times 10^{-16}$. (b) Setting $\sigma= 1.1\times 10^{-16}$ and varying $\theta$. The dotted curve presents a critical temperature $T_0 \approx 5.85\times 10^{9}$ while the dashed line presents $T_0 \approx 3.55\times 10^{9}$. (b) Setting $\theta= 2.4\times 10^{-14}$ and varying $\sigma$. The solid, dashed and dotted lines have  $T_0 \approx 5.1\times 10^{9}$, $ T_0 \approx 5.46 \times 10^{9}$ and $T_0 \approx 5.84 \times 10^{9}$, respectively.}
	\label{fig:charges}
\end{figure}

Taking into account the possibility of boson-antiboson pair creation, as presented in the section \ref{BLSM}, we fixed the charge $Q$ so that the transition temperature $T_{0}$ can be obtained numerically by equation (\ref{rhop}). The results for ratio of charges in the ground state as a function of the temperature is presented in Figure [\ref{fig:charges}], where one can be seen that non-commutative effects appears at lower temperatures than for the system without antibosons. It is interesting the way the critical temperature varies according to the choice of the parameters $\theta$ and $\sigma$. An increase in $\theta$ causes a decrease in the transition temperature, but when $\theta$ approaches zero or $\sigma$ is large enough the non-commutative curve coincides with the commutative curve what is in accord to the regularization given by a Gaussian-type distribution in Eq.~(\ref{delta}).

\begin{figure*}[!h]
	\centering
		\includegraphics[width=10.0cm,height=6.6cm]{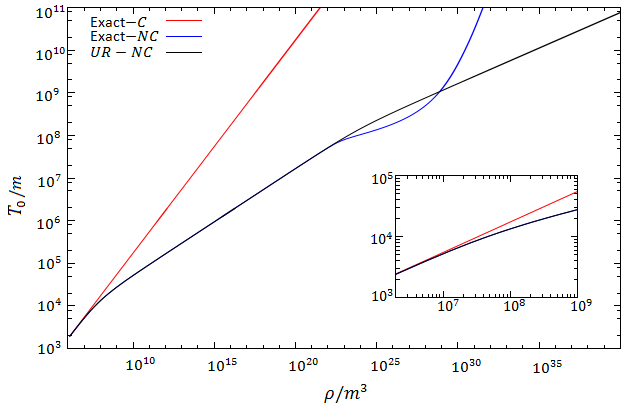}
 \caption[Critical temperatures $T_0$ as a function of charge density $\rho$.]{Numeric results of $T_0$ as a function of charge density $\rho$, comparing commutative (exact-C), non-commutative (exact-NC) and non-commutative ultra-relativistic (UR-NC) cases. The parameters are $\theta=2.4\times 10^{-14}$, $\sigma=1.1\times 10^{-16}$ and $m=5.1\times 10^{5}$.}
	\label{fig:TcXrho}
\end{figure*}

\begin{figure*}[!h]
	\centering
		\includegraphics[width=10.0cm,height=7.6cm]{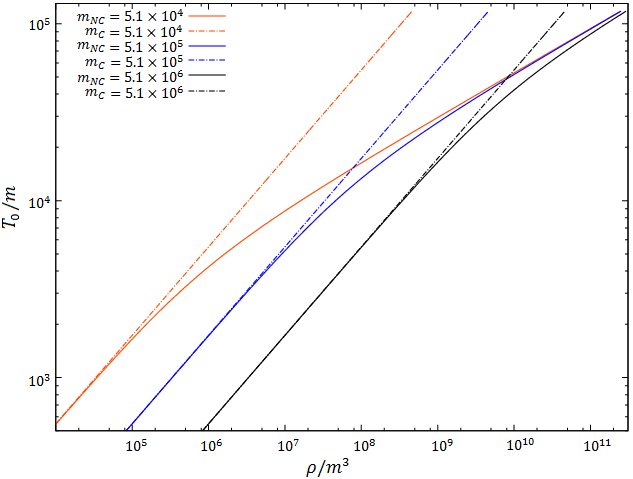}
	\caption[Numeric results of $T_0$ as a function of charge density $\rho$ for different mass.]{ Numeric results of $T_0$ as a function of charge density $\rho$ for bosons with various values
of mass $m=5.1\times 10^{4}$, $m=5.1\times 10^{5}$, $m=5.1\times 10^{6}$. The solids lines represent the non-commutative situation and the dashed lines indicate the commutative case.}
	\label{fig:X}
\end{figure*}

The results of $T_0$ for different values of charge density can be seen in Figure [\ref{fig:TcXrho}] comparing the numerical procedure with  UR-NC regime, which shows the efficiency region of this approximation  based on the expansion of the parameters $\theta$ and $\sigma$ in determining the thermodynamic quantities of interest. 
Another interesting aspect presented by this new non-commutative BEC arises in the curves of  transition temperatures versus charge density for different values of mass shown in Figure [\ref{fig:X}].

\begin{figure}[!h]
	\subfigure[\label{fig:ene}]{\includegraphics[width=8.5cm,height=6.5cm]{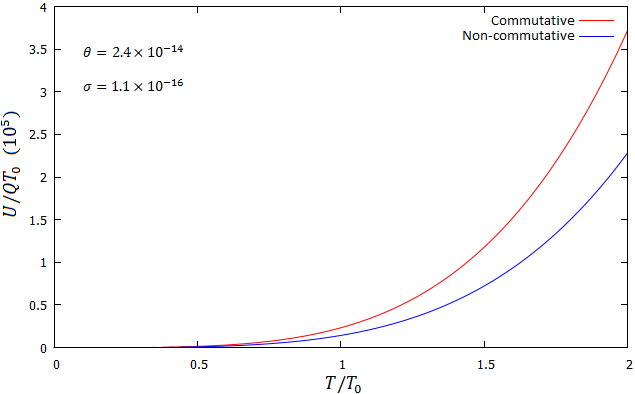}}
	\subfigure[\label{fig:press}]{\includegraphics[width=8.5cm,height=6.5cm]{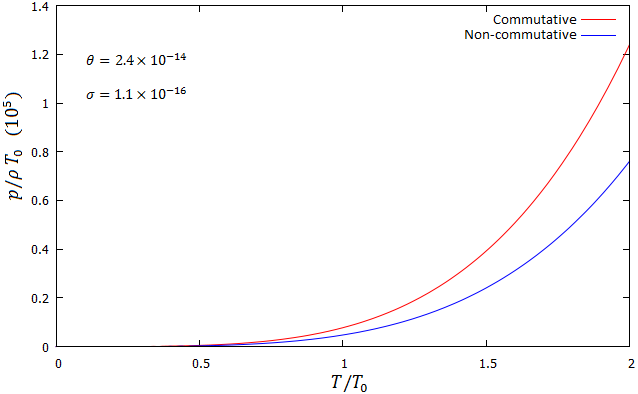}} \\
	\subfigure[\label{fig:cvQ}]{\includegraphics[width=8.5cm,height=6.5cm]{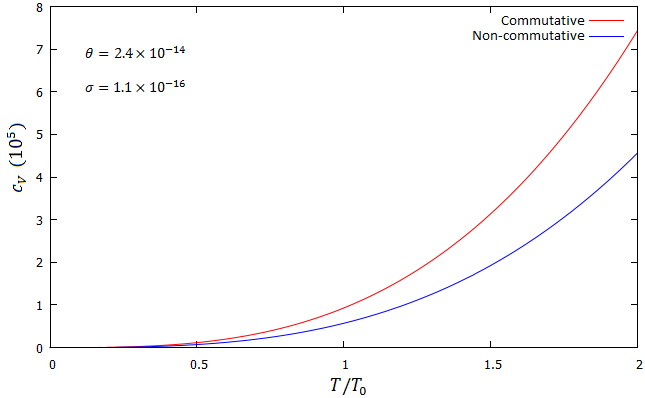}}
 \caption{Comparing the commutative and non-commutative cases with a net fixed charge density $\rho= 6.12\times 10^{24}$, setting $\theta=2.4\times 10^{-14}$ and $\sigma=1.1\times 10^{-16}$. (a)  Internal energy per particle as a function of the temperature. (b) Pressure per density of particles as a function of the temperature. (c) Specific heat as a function of temperature.}
\label{fig:enepres}
\end{figure}

The non-commutative effects on internal energy, pressure and specific heat are shown in  Figure [\ref{fig:enepres}] together with those obtained through the usual commutative theory. In this regime of temperature, the numerical results obtained using the equations (\ref{energia2}), (\ref{pressao2}) and  (\ref{diffs}) are the same as those obtained by the ultra-relativistic regime. We can  observe that the specific heat has no gap for a fixed net charge $Q$. This is because the mechanism of pair creation of particles and antiparticles removes this anomaly in three spatial dimensions \cite{frota}.
%

%
\vspace*{4.5cm}
\section{Conclusions}
\label{chp:4}

We investigated the thermodynamic properties of a relativistic boson gas in the scenario of quantum field theory based on non-commutative fields. We aimed to obtain the thermodynamic equations that describe a Bose-Einstein condensate, such as temperature transition, fraction condensate, internal energy, pressure and specific heat, in order to investigate the corrections  caused in the system due to the introduction of the non-commutativity.

The first consequence of the non-commutativity appears in the dispersion relation, which leads to modifications of the  thermodynamic quantities. To understand these modifications and their dependence on the  non-commutative parameters ($\theta$ and $\sigma$) introduced in the theory we mainly study the graphics obtained numerically and by approximated methods at some regimes. These modified thermodynamic quantities are then compared to known results of the commutative theory in the literature. Specially, in the case without  antibosons,  the specific heat around the critical temperature has a larger gap in the non-commutative case. In low temperature physics this effect can be even more interesting in high-$T_c$ superconductors. The specific heat also presents a lump for temperatures immediately larger than the critical temperature which are very similar to the thermodynamic anomaly of water.

The analysis of these results shows that the non-commutative effects cause significant contributions in high critical temperatures, what is related to the fact that the non-commutativity  is an effect only perceptible at the Planck scale, because the non-commutative parameters are expected to be very small in the context of quantized spacetime. The system with antibosons has non-commutative effects at relatively lower critical temperatures compared to the system without antibosons. In this case the specific heat has no gap for a fixed charge $Q$. This is because the mechanism of pair creation of particles removes this anomaly in three spatial dimensions. In this case one can recover the well-known Haber and Weldon critical temperature in the limit as $\theta\to0$. On the other hand, in the limit of massless non-commutative scalar fields one can still find a finite critical temperature depending on $\theta$ for the non-commutative Bose-Einstein condensate.

Among the prospects to apply this work is to explore the Bose-Einstein condensation with non-commutative scalar fields in curved spaces.


\acknowledgments

We would like to thank J.B. Silva, C. Furtado, L. Barosi and A.R. Queiroz for invaluable discussions. We also thank CNPq and CAPES  for partial financial support.
\newpage
\appendix
\section{ \label{appendixA} The commutation relations}

In this appendix we calculate the commutation relations that appear in the section (\ref{EANC}). Using the equations (\ref{eq1NC}) we found
\begin{center}
\begin{eqnarray}
\label{Ap1}
 \left[ \hat{\varphi}^{a}(\vec{x},t),\hat{\varphi}^{b}(\vec{y},t)\right]&=& \left[ \varphi^{a}(\vec{x},t)-\frac{1}{2}\epsilon^{ac}\theta\pi_{c}(\vec{x},t), \varphi^{b}(\vec{y},t)-\frac{1}{2}\epsilon^{bd}\theta\pi_{d}(\vec{y},t)\right] \nonumber\\ 
&=& \left[ \varphi^{a}(\vec{x},t), \varphi^{b}(\vec{y},t)\right] -\frac{1}{2}\epsilon^{bd}\theta\left[ \varphi^{a}(\vec{x},t),\pi_{d}(\vec{y},t)\right] \nonumber \\
& &-\frac{1}{2}\epsilon^{ac}\theta\left[\pi_{c}(\vec{x},t), \varphi^{b}(\vec{y},t)\right] +\frac{1}{4}\epsilon^{ac}\epsilon^{bd}\theta^2\left[\pi_{c}(\vec{x},t), \pi_{d}(\vec{y},t)\right] \nonumber\\
&=&-\frac{i}{2}\epsilon^{bd}\theta\delta^{a}_{d}\delta(\vec{x}-\vec{y})+\frac{i}{2}\epsilon^{ac}\theta\delta^{b}_{c}\delta(\vec{x}-\vec{y}) \nonumber \\
&=&i\epsilon^{ab}\theta\delta(\vec{x}-\vec{y}), \\ \nonumber \\
 \left[ \hat{\varphi}^{a}(\vec{x},t),\hat{\pi}_{b}(\vec{y},t)\right]&=&\left[\varphi^{a}(\vec{x},t)-\frac{1}{2}\epsilon^{ac}\theta\pi_{c}(\vec{x},t) ,\pi_{b}(\vec{y},t)\right] \nonumber \\
&=& \left[\varphi^{a}(\vec{x},t),\pi_{b}(\vec{y},t)\right] -\frac{1}{2}\epsilon^{ac}\theta
\left[\pi_{c}(\vec{x},t) ,\pi_{b}(\vec{y},t) \right]\nonumber \\
&=& i\delta_{b}^{a}\delta(\vec{x}-\vec{y}), \\ \nonumber \\
\left[ \hat{\pi}_{a}(\vec{x},t),\hat{\pi}_{b}(\vec{y},t)\right]&=&\left[\pi_{a}(\vec{x},t),\pi_{b}(\vec{y},t)\right]=0.
\end{eqnarray}
\end{center}
The commutation relations components for the Fourier transformations are obtained using (\ref{eq2NC})
\begin{center}
\begin{eqnarray}
\left[\hat{\varphi}^{a}_{\vec{n}}\,,\,\hat{\varphi}^{b}_{\vec{m}}\right] &=& \left[\varphi^{a}_{\vec{n}}-\frac{1}{2R^3}\epsilon^{ac}\theta(n)\pi^{c}_{-\vec{n}}\,,\, \varphi^{b}_{\vec{m}}-\frac{1}{2R^3}\epsilon^{bd}\theta(m)\pi^{d}_{-\vec{m}} \right] \nonumber \\
&=& \left[\varphi^{a}_{\vec{n}},\varphi^{b}_{\vec{m}}\right]-\frac{1}{2R^3}\epsilon^{bd}\theta(m)\left[\varphi^{a}_{\vec{n}}, \pi^{d}_{-\vec{m}}\right] -\frac{1}{2R^3}\epsilon^{ac}\theta(n)\left[\pi^{c}_{-\vec{n}},\varphi^{b}_{\vec{m}}\right] \nonumber \\
& &+ \frac{1}{4R^6}\epsilon^{ac}\epsilon^{bd}\theta(n)\theta(m) \left[\pi^{c}_{-\vec{n}},\pi^{d}_{-\vec{m}}\right] \nonumber\\
&=&-\frac{i}{2R^3}\epsilon^{bd}\theta(m)\delta^{ad}\delta_{\vec{n},-\vec{m}} +\frac{i}{2R^3}\epsilon^{ac}\theta(n)\delta^{cb}\delta_{-\vec{n},\vec{m}}\nonumber\\
&=&\frac{i\epsilon^{ab}\theta(n)}{R^3}\delta_{\vec{n}+\vec{m},0}, \\ \nonumber \\
\left[\hat{\varphi}^{a}_{\vec{n}},\hat{\pi}^{b}_{\vec{m}}\right] &=& \left[\varphi^{a}_{\vec{n}}-\frac{1}{2R^3}\epsilon^{ac}\theta(n)\pi^{c}_{-\vec{n}}\,,\,\pi^{b}_{\vec{m}}\right]\nonumber \\
&=&\left[\varphi^{a}_{\vec{n}},\pi^{b}_{\vec{m}}\right] -\frac{1}{2R^3}\epsilon^{ac}\theta(n)\left[\pi^{c}_{-\vec{n}}\,,\,\pi^{b}_{\vec{m}}\right]\nonumber \\
&=&i\delta^{ab}\delta_{\vec{n},\vec{m}},\\ \nonumber \\
\left[\hat{\pi}^{a}_{\vec{n}},\hat{\pi}^{b}_{\vec{m}}\right] &=&\left[\pi^{a}_{\vec{n}},\pi^{b}_{\vec{m}}\right] =0.
\end{eqnarray}
\end{center}
To obtain the commutation relations for the operators ${a_{\vec{n}}^{i}}^{\dagger}$ and $a_{\vec{n}}^{i}$ we use equations (\ref{CA}) and (\ref{CAA}) 
\begin{center}
\begin{eqnarray}
\left[a^{i}_{\vec{m}},a^{j}_{\vec{n}}\right] &=& \left[\sqrt{\frac{\Delta_{\vec{m}}}{2}}\left( \varphi^{i}_{\vec{m}} + i\frac{\pi^{i}_{-\vec{m}}}{\Delta_{\vec{m}}} \right)\,,\, \sqrt{\frac{\Delta_{\vec{n}}}{2}}\left( \varphi^{j}_{\vec{n}} + i\frac{\pi^{j}_{-\vec{n}}}{\Delta_{\vec{n}}} \right)\right]\nonumber \\
&=&\frac{\sqrt{\Delta_{\vec{m}}\Delta_{\vec{n}}}}{2}\left(\left[\varphi^{i}_{\vec{m}},\varphi^{j}_{\vec{n}}\right]+ \frac{i}{\Delta_{\vec{n}}} \left[\varphi^{i}_{\vec{m}},\pi^{j}_{-\vec{n}}\right]+  \frac{i}{\Delta_{\vec{m}}} \left[\pi^{i}_{-\vec{m}},\varphi^{j}_{\vec{n}}\right]- \frac{1}{\Delta_{\vec{m}}\Delta_{\vec{n}}}\left[\pi^{i}_{-\vec{m}},\pi^{j}_{-\vec{n}}\right]\right)\nonumber \\
&=& \frac{\sqrt{\Delta_{\vec{m}}\Delta_{\vec{n}}}}{2}\left(-\frac{1}{\Delta_{\vec{n}}}\delta^{ij}\delta_{\vec{m}+\vec{n},0} + \frac{1}{\Delta_{\vec{m}}}\delta^{ij}\delta_{\vec{m}+\vec{n},0} \right)\nonumber\\
&=&0,\\ \nonumber \\
\left[{a^{i}_{\vec{m}}}^\dagger,{a^{j}_{\vec{n}}}^\dagger\right] &=& \left[\sqrt{\frac{\Delta_{\vec{m}}}{2}}\left( \varphi^{i}_{-\vec{m}} - i\frac{\pi^{i}_{\vec{m}}}{\Delta_{\vec{m}}} \right)\,,\, \sqrt{\frac{\Delta_{\vec{n}}}{2}}\left( \varphi^{j}_{-\vec{n}} - i\frac{\pi^{j}_{\vec{n}}}{\Delta_{\vec{n}}} \right)\right]\nonumber \\
&=&\frac{\sqrt{\Delta_{\vec{m}}\Delta_{\vec{n}}}}{2}\left(\left[\varphi^{i}_{-\vec{m}},\varphi^{j}_{\vec{-n}}\right]- \frac{i}{\Delta_{\vec{n}}} \left[\varphi^{i}_{-\vec{m}},\pi^{j}_{\vec{n}}\right]-  \frac{i}{\Delta_{\vec{m}}} \left[\pi^{i}_{\vec{m}},\varphi^{j}_{-\vec{n}}\right]- \frac{1}{\Delta_{\vec{m}}\Delta_{\vec{n}}}\left[\pi^{i}_{\vec{m}},\pi^{j}_{\vec{n}}\right]\right)\nonumber \\
&=& \frac{\sqrt{\Delta_{\vec{m}}\Delta_{\vec{n}}}}{2}\left(\frac{1}{\Delta_{\vec{n}}}\delta^{ij}\delta_{\vec{m}+\vec{n},0} - \frac{1}{\Delta_{\vec{m}}}\delta^{ij}\delta_{\vec{m}+\vec{n},0} \right)\nonumber\\
&=&0,
\end{eqnarray}
\end{center}
\begin{center}
\begin{eqnarray}
\left[a^{i}_{\vec{m}},{a^{j}_{\vec{n}}}^\dagger\right] &=& \left[\sqrt{\frac{\Delta_{\vec{m}}}{2}}\left( \varphi^{i}_{\vec{m}} + i\frac{\pi^{i}_{-\vec{m}}}{\Delta_{\vec{m}}} \right)\,,\, \sqrt{\frac{\Delta_{\vec{n}}}{2}}\left( \varphi^{j}_{-\vec{n}} - i\frac{\pi^{j}_{\vec{n}}}{\Delta_{\vec{n}}} \right)\right]\nonumber \\
&=&\frac{\sqrt{\Delta_{\vec{m}}\Delta_{\vec{n}}}}{2}\left(\left[\varphi^{i}_{\vec{m}},\varphi^{j}_{\vec{-n}}\right]- \frac{i}{\Delta_{\vec{n}}} \left[\varphi^{i}_{\vec{m}},\pi^{j}_{\vec{n}}\right]+  \frac{i}{\Delta_{\vec{m}}} \left[\pi^{i}_{-\vec{m}},\varphi^{j}_{-\vec{n}}\right]+ \frac{1}{\Delta_{\vec{m}}\Delta_{\vec{n}}}\left[\pi^{i}_{-\vec{m}},\pi^{j}_{\vec{n}}\right]\right)\nonumber \\
&=& \frac{\sqrt{\Delta_{\vec{m}}\Delta_{\vec{n}}}}{2}\left(\frac{1}{\Delta_{\vec{n}}}\delta^{ij}\delta_{\vec{m},\vec{n}} + \frac{1}{\Delta_{\vec{m}}}\delta^{ij}\delta_{\vec{m},\vec{n}} \right)\nonumber\\
&=&\delta^{ij}\delta_{\vec{m},\vec{n}}.
\end{eqnarray}
\end{center}
In the following we calculate the commutation relations (\ref{comutador4})  using the expressions (\ref{oper1}), then
\begin{center}
\begin{eqnarray}
\left[A^{1}_{\vec{m}},{A^{1}_{\vec{n}}}^\dagger\right]&=&\frac{1}{2}\left[a^{1}_{\vec{m}} - i a^{2}_{\vec{m}}, {a^{1}_{\vec{n}}}^\dagger + i {a^{2}_{\vec{n}}}^\dagger \right] \nonumber \\
&=&\frac{1}{2} \left([a^{1}_{\vec{m}}, {a^{1}_{\vec{n}}}^\dagger] + i[a^{1}_{\vec{m}}, {a^{2}_{\vec{n}}}^\dagger] -i[a^{2}_{\vec{m}}, {a^{1}_{\vec{n}}}^\dagger] +[a^{2}_{\vec{m}}, {a^{2}_{\vec{n}}}^\dagger]\right)\nonumber \\
&=& \delta_{\vec{m},\vec{n}}, \\ \nonumber \\
\left[A^{2}_{\vec{m}},{A^{2}_{\vec{n}}}^\dagger\right]&=&\frac{1}{2}\left[a^{1}_{\vec{m}} + i a^{2}_{\vec{m}}, {a^{1}_{\vec{n}}}^\dagger - i {a^{2}_{\vec{n}}}^\dagger \right] \nonumber \\
&=&\frac{1}{2} \left([a^{1}_{\vec{m}}, {a^{1}_{\vec{n}}}^\dagger] - i[a^{1}_{\vec{m}}, {a^{2}_{\vec{n}}}^\dagger] +i[a^{2}_{\vec{m}}, {a^{1}_{\vec{n}}}^\dagger] +[a^{2}_{\vec{m}}, {a^{2}_{\vec{n}}}^\dagger]\right)\nonumber \\
&=& \delta_{\vec{m},\vec{n}},
\end{eqnarray}
\end{center}
and
\begin{center}
\begin{eqnarray}
\left[A^{1}_{\vec{m}},{A^{2}_{\vec{n}}}^\dagger\right]&=&\frac{1}{2}\left[a^{1}_{\vec{m}} - i a^{2}_{\vec{m}}, {a^{1}_{\vec{n}}}^\dagger - i {a^{2}_{\vec{n}}}^\dagger \right] \nonumber \\
&=&\frac{1}{2} \left([a^{1}_{\vec{m}}, {a^{1}_{\vec{n}}}^\dagger] - i[a^{1}_{\vec{m}}, {a^{2}_{\vec{n}}}^\dagger] - i[a^{2}_{\vec{m}}, {a^{1}_{\vec{n}}}^\dagger] - [a^{2}_{\vec{m}}, {a^{2}_{\vec{n}}}^\dagger]\right)\nonumber \\
&=& 0, \\ \nonumber \\
\left[A^{2}_{\vec{m}},{A^{1}_{\vec{n}}}^\dagger\right]&=&\frac{1}{2}\left[a^{1}_{\vec{m}} + i a^{2}_{\vec{m}}, {a^{1}_{\vec{n}}}^\dagger + i {a^{2}_{\vec{n}}}^\dagger \right] \nonumber \\
&=&\frac{1}{2} \left([a^{1}_{\vec{m}}, {a^{1}_{\vec{n}}}^\dagger] + i[a^{1}_{\vec{m}}, {a^{2}_{\vec{n}}}^\dagger] + i[a^{2}_{\vec{m}}, {a^{1}_{\vec{n}}}^\dagger] - [a^{2}_{\vec{m}}, {a^{2}_{\vec{n}}}^\dagger]\right)\nonumber \\
&=& 0.
\end{eqnarray}
\end{center}
Therefore,
\begin{center}
\begin{eqnarray}
\left[A^{i}_{\vec{m}},{A^{j}_{\vec{n}}}^\dagger\right] = \delta^{ij}\delta_{\vec{m},\vec{n}}.
\end{eqnarray}
\end{center}

\newpage


\end{document}